\documentclass[11pt]{article}
\usepackage{epsfig}
\usepackage{amsmath}
\usepackage{amsfonts}
\usepackage{amssymb}

\numberwithin{equation}{section}

\title{Solutions of the WDVV Equations and Integrable Hierarchies of KP type}
\author{Henrik Aratyn\\
\\
Department of Physics,\\
University of Illinois at Chicago,\\
845 W. Taylor St.,\\
Chicago, IL 60607-7059\\
e-mail: aratyn@uic.edu\\
\and Johan van de Leur\\
\\
Mathematical Institute,\\
University of Utrecht,\\
P.O. Box 80010, 3508 TA Utrecht,\\
The Netherlands\\
e-mail: vdleur@math.uu.nl}

\newtheorem{corollary}{Corollary}[section]
\newtheorem{definition}{Definition}[section]
\newtheorem{lemma}{Lemma}[section]
\newtheorem{proposition}{Proposition}[section]
\newtheorem{theorem}{Theorem}[section]

\newcommand\threemat[9]{\left(\begin{array}{ccc}  %%   3x3 matrix  
{#1} & {#2} & {#3}\\ {#4} & {#5} & {#6}\\
{#7} & {#8} & {#9} \end{array} \right)}

\begin{document}

\maketitle

\begin{abstract}
We show that reductions of KP hierarchies related to the loop 
algebra of $SL_n$ with homogeneous gradation
give solutions of the Darboux-Egoroff system of PDE's. 
Using explicit dressing matrices of the Riemann-Hilbert problem 
generalized to include a set of commuting additional 
symmetries, we construct solutions of
the Witten--Dijkgraaf--E. Verlinde--H. Verlinde equations.
\end{abstract}

\section{Introduction}

This paper deals with symmetries of integrable models and their
connection to the topological field theory and Frobenius manifolds.
Our results establish a new and fundamental link between
Darboux-Egoroff metric systems and additional symmetry-flows of the
integrable hierarchies of KP type. This connection goes beyond a 
dispersionless limit.

The description of $n$-orthogonal curvilinear coordinate systems in $\mathbb{R}^n$ was 
one of the classical puzzles of the 19th century.
Well-known mathematicians such as Gauss, Lam\'e, Cayley and Darboux have contributed to its solution. 
The problem is formulated as follows. Find all coordinate systems 
\begin{equation}
\label{D0}
\begin{aligned}
u_i=u_i(x^1,x^2,\ldots,x^n),\\
\det\left(\left(\frac{\partial u_i}{\partial x^j}\right)_{1\le i,j\le n}\right)
\ne 0,
\end{aligned}
\end{equation} 
satisfying the orthogonality condition
\[
\sum_{k=1}^n \frac{\partial u_i}{\partial x^k}
\frac{\partial u_j}{\partial x^i}=0,\quad \text{for } i\ne j.
\]
It was Darboux who gathered most results on this topic and published them  in 1910 in his 
{\it Le\c cons sur les syst\`emes orthogonaux et les coordonn\'ees curvilignes} \cite{Dar}. 
For more historical details see the paper \cite{Zak} by Zakharov.

The metric tensor in $\mathbb{R}^n$ in the coordinate system $u_i$ is diagonal:
\begin{equation}
\label{D1}
ds^2=\sum_{i=1}^n h^2_i(u) (du_i)^2,\qquad u=(u_1,\dots,u_n),
\end{equation}
where
\begin{equation}
\label{D2}
h_i^2(u)=\sum_{k=1}^n \left(\frac{\partial x^i}{\partial u_k}\right)^2.
\end{equation}
The flatness of the metric imposes a condition on the Lam\'e 
coefficients $h_i$  of the following form 
\begin{equation}
\label{D3}
\begin{aligned}
\partial_k\beta_{ij}(u)=\beta_{ik}(u)\beta_{kj}(u),\quad &i\ne k\ne j,\\
\frac{\partial\beta_{ij}(u)}{\partial u_i}+\frac{\partial\beta_{ji}(u)}{\partial u_j}
+\sum_{k\ne i,j}\beta_{ki}\beta_{kj}=0,\quad &i\ne j,
\end{aligned}
\end{equation}
here this is expressed in Darboux' rotation coefficients
\begin{equation}
\label{D4}
\beta_{ij}(u)=\frac{1}{h_i(u)}\frac{\partial h_j(u)}{\partial u_i}.
\end{equation}
For rotation coefficients $\beta_{ij}$ that satisfy
\begin{equation}
\label{D5}
\beta_{ij}(u)=\beta_{ji}(u),
\end{equation}
the corresponding metric is called an Egoroff metric, 
in honour of D.F. Egoroff who gave its 
complete description  \cite{Ego}.
 
B. Dubrovin noticed \cite{Du3a}, \cite{Du1} that the equations (\ref{D3}) and (\ref{D5}), 
which he called the Darboux-Egoroff system,  also describe the local classification
of massive topological field theories. 
An important role in physics literature on two
dimensional field theory is played by a remarkably system of partial
differential equations commonly known as the
Witten-Dijkgraaf-Verlinde-Verlinde (WDVV) equations \cite{DVV}, \cite{W1}. These equations
determine deformations of 2-dimensional topological field theories.
Dubrovin \cite{Du3a},\cite{Du1} connected the Frobenius structure to
any solution of the WDVV-equation thus 
providing a ``coordinate-free'' approach
to the topological field theory. Frobenius manifolds are objects from 
the differential geometry which arise in a number of
different areas of mathematics, such as quantum cohomology, 
Gromov-Witten invariants, theory of
singularities, Hurwitz spaces, Coxeter groups.
One of the most interesting connections from a physics point of view is a 
link to the integrable systems.

We obtain a class of integrable hierarchies which provides a setting for
solutions to the Darboux-Egoroff metric systems.
The relevant integrable models turn out to be connected with generalizations
of the integrable structure related to the Nonlinear Schr\"{o}dinger equation.
In the pseudo-differential calculus framework of the Sato theory the relevant models
originate from the Lax structure given by
\begin{equation}
{\cal L} = {\partial_x} + \sum_{i=1}^{m} \Phi_i 
{\partial_x}^{-1} \Psi_i \, .
\label{lax}
\end{equation}
The corresponding Baker-Akhiezer (BA) wave function $\psi_{BA}$ 
which enters the linear spectral problem  ${\cal L}  \psi_{BA} = \lambda \psi_{BA}$ 
is given as  
\begin{equation}
\psi_{BA} (t, \lambda) = \frac{ \tau (t-[\lambda^{-1}]) 
 }{ \tau (t)}  e^{ \sum_{j=1}^{\infty} \lambda^{j} t_j} \, , 
\label{ba-fct}
\end{equation}
in terms of the $\tau$-function such that $ \partial _x \partial   \ln \tau /\partial t_n
= {\rm Res}  ({\cal L}^n)$.
In (\ref{ba-fct}) we used the multi-time notation $(t-[\lambda^{-1}]) = (t_1 - 1/\lambda, t_2 -
1/2\lambda^2, {\ldots} )$ 

For the Lax operator in (\ref{lax}) the action of the isospectral flows
on the Lax operator ${\cal L} $ is given by
\begin{equation}
\frac{\partial }{ \partial t_n} {\cal L}  = \lbrack ({\cal
L}^n)_{+}\, , \,  {\cal L}  \rbrack \, ,
\label{iso-lax}
\end{equation}
where as usual $({\cal
L}^n)_{+}$ stands for the differential part of the pseudo-differential operator.
%In partcular, it follows that $\frac{\partial }{ \partial t_1}= \partial_x$ and we therefore
%make identification $t_1=x$.

The isospectral flows from (\ref{iso-lax}) induce the following flows on the 
eigenfunctions $\Phi_i$ and adjoint eigenfunctions $\Psi_i$
with $i=1,{\ldots} ,m$ :
\begin{equation}
\frac{\partial }{ \partial t_n} \Phi_i = ({\cal
L}^n)_{+} (\Phi_i) \quad ; \quad \frac{\partial }{ \partial t_n} \Psi_i = - ({\cal
L}^{n})^*_{+} (\Psi_i)  \, .  
\label{iso-pp}
\end{equation}

These models possess a rich structure of additional, with respect to the
isospectral flows, symmetries.
These symmetry flows will be studied within the pseudo-differential calculus framework 
where they are generated by the squared
eigenfunction potentials \cite{sep} and within the algebraic formalism
obtained by extending the standard Riemann-Hilbert problem.
In the latter formalism the relevant
systems are characterized through their Lie algebraic context as
integrable models arising through generalized Drinfeld-Sokolov scheme from
the $sl(n)$ algebras with the homogeneous gradation.

After further reduction to CKP sub-hierarchy \cite{Date81}, which brings conditions
$\Phi_i=\Psi_i$, we obtain a model whose abelian subalgebra of additional symmetries 
provide canonical coordinates of the related Darboux-Egoroff systems.
This connection will be used to 
explicitly construct solutions to the Darboux-Egoroff system
and the Witten-Dijkgraaf-Verlinde-Verlinde equations in the topological field theory.

In a separate publication we will address the issue of how several
structures of the underlying integrable models (including Virasoro symmetry)
carry over to the Darboux-Egoroff system.

In Section 2, we discuss the Darboux-Egoroff metrics and the
underlying linear spectral system from which the Darboux-Egoroff system
can be obtained via compatibility equations.
Section 3 defines the constrained integrable KP hierarchy, which is a subject
of our study, in terms of the Sato Grassmannian and within the framework
of the pseudo-differential operator calculus.
In Section 4, the Riemann-Hilbert problem is extended to incorporate
the  constrained integrable KP hierarchy  augmented by additional symmetry
flows. We succeed, in this section, to find an explicit expression for the 
dressing matrix allowing us to solve the extended Riemann-Hilbert problem 
in terms of the $\tau$-function and the eigenfunctions $\Phi_i$ and the 
adjoint eigenfunctions $\Psi_i$.
Section 5 introduces the CKP reduction of the constrained integrable 
KP hierarchy imposing equality of the eigenfunctions $\Phi_i$ and the adjoint 
eigenfunctions $\Psi_i$.
Finally, in Section 6, we reproduce the fundamental objects defining 
the Darboux-Egoroff system presented in Section 2 in terms of the 
integrable structure derived in the previous sections.

\section{The Darboux-Egoroff system}
\label{Darb}

We now describe in some more details a relation
between the Darboux-Egoroff system (\ref{D3}), (\ref{D5}) and the local classification
of massive topological field theories \cite{Du3a}, \cite{Du1}.
First of all, the flat coordinates $x^1,\dots,x^n$ of the Darboux-Egoroff 
metric (\ref{D1}) can be found from the linear system
\begin{equation}
\label{D6}
\frac{\partial^2 x^k}{\partial u_i\partial u_j}=\Gamma_{ij}^i \frac{\partial x^k}
{\partial u_i} 
+ \Gamma_{ji}^j \frac{\partial x^k}{\partial u_j},
\quad i\ne j;\qquad\quad
\frac{\partial^2 x^k}{\partial u_i^2 }=\sum_{j=1}^n \Gamma_{ii}^j \frac{\partial x^k}{\partial u_j},
\end{equation}
where $\Gamma_{ij}^k$ are the Christoffel symbols of the Levi-Civita connection:
\begin{equation}
\label{f09}
\Gamma_{ij}^i=\frac{1}{h_i}\frac{\partial h_i}{\partial u_j},\qquad
\Gamma_{ii}^j=(2\delta_{ij}-1)\frac{h_i}{h^2_j}\frac{\partial h_i}{\partial u_j}.
\end{equation}
For the flat coordinates $x^i$, $1\le i\le n$, of the metric (\ref{D1}), the
functions
\begin{equation}
\label{D7}
c_{k\ell}^m(x) = \sum_{i=1}^n \frac{\partial x^m}{\partial u_i} \frac{\partial
u_i}{\partial x^k}
\frac{\partial u_i}{\partial x^\ell},
\end{equation}
form  the structure constants of the commutative  algebra of the primary fields $\phi_i$ of a  
topological field theory, i.e.
\[
\phi_k\phi_\ell=\sum_m c_{k\ell}^m \phi_m.
\]
Associativity of this algebra imposes on the structure constants 
a relation :
\begin{equation}
\label{D8}
\sum_{k=1}^n c_{ij}^k(x) c_{km}^\ell(x) = \sum_{k=1}^nc_{jm}^k(x)
c_{ik}^\ell(x) .
\end{equation}
If one writes down these equations for the function $F(x)$ for which
\begin{equation}\label{D9}
\frac{\partial^3 F(x)}{\partial x^k \partial x^\ell \partial x^m}=c_{k\ell
m}(x)=\sum_{i=1}^n\eta_{mi} c_{k\ell}^i(x),\quad
\mbox{where}\quad \eta_{pq}=\sum_{i=1}^n h_i^2(u) \frac{\partial u_i}{\partial
x^p}
\frac{\partial u_i}{\partial x^q},
\end{equation}
with the constraint
\[
\frac{\partial^3 F(x)}{\partial x^1 \partial x^\ell \partial x^m}=\eta_{\ell
m},
\]
one obtains the well-known  Witten-Dijkgraaf-E. Verlinde-H. Verlinde
(WDVV)-equations \cite{W1}, \cite{DVV} for the prepotential $F(x)$.
In fact, Dubrovin showed \cite{Du3a} that one can find these $c_{k\ell
m}(x)$ and  $\eta_{pq}$ of (\ref{D9}) as follows. The Darboux-Egoroff system 
can be represented as the compatibility equations of the following linear system depending on a spectral parameter $\lambda$:
\begin{equation}
\label{D10}
\begin{aligned}
\frac{\partial \psi_{ik}(u,\lambda)}{\partial u_j}&=\beta_{ij}(u)\psi_{jk}(u,\lambda),\quad i\ne j\\
\sum_{k=1}^n \frac{\partial \psi_{ij}(u,\lambda)}{\partial u_k}&=\lambda\psi_{ij}(u,\lambda),
\end{aligned}
\end{equation}
Solving this system for $\lambda=0$ , i.e., finding $\psi_{ij}(u)$ that satisfy
\begin{equation}
\label{D11}
\begin{aligned}
\frac{\partial \psi_{ik}(u)}{\partial u_j}&=\beta_{ij}(u)\psi_{jk}(u),\quad i\ne j\\
\sum_{k=1}^n \frac{\partial \psi_{ij}(u)}{\partial u_k}&=0,
\end{aligned}
\end{equation}
for a given solution $\beta_{ij}(u)$ of the Darboux-Egoroff system, leads to "a local 
classification of complex semisimple Frobenius manifolds":
\begin{proposition}
\label{Prop1}\cite{Du3a}
On the domain
$u_i\ne u_j$ and $\psi_{11}\psi_{21}\cdots\psi_{n1}\ne 0$, one has
\begin{equation}
\label{D12}
\begin{aligned}
h_i(u)&=\psi_{i1}(u),\\
\eta_{\alpha\beta}&=\sum_{i=1}^n
\psi_{i\alpha}(u)\psi_{i\beta}(u),\\
\sum_\beta \eta_{\alpha\beta}\frac{\partial x^\beta(u)}{\partial u_i}&=\psi_{i1}(u)\psi_{i\alpha}(u),\\
c_{\alpha\beta\gamma}(x(u))&=\sum_{i=1}^n
\frac{\psi_{i\alpha}(u)\psi_{i\beta}(u)\psi_{i\gamma}(u)}{\psi_{i1}(u)}.
\end{aligned}
\end{equation}
\end{proposition}
We refer the reader for the definition of Frobenius manifold to \cite{Du3a} or \cite{Du1}. 
This definition is different from the one given in e.g \cite{Du2}.
We do not assume the quasi-homogeneity condition here.
Note that this system (\ref{D12}) is not unique. For any given solution $\beta_{ij}(u)$ 
of the Darboux-Egoroff system, there exists an $n$-parameter family of Lam\'e coefficients and of Egoroff metrics.

Assume for simplicity that $\eta_{\alpha\beta}=\delta_{\alpha\beta}$,
so $c_{k\ell m}=c_{k\ell }^m$.
Following Akhmetshin, Krichever and Volvovski \cite{AKrV} (see also \cite{Kr} and \cite{vdL1}),
we can even construct the prepotential $F(x)$.
There exist $n$ unique solutions $\psi_{ij}(u,\lambda)$, $1\le j\le n$, of (\ref{D10}) with initial conditions
$\psi_{ij}(0,\lambda)=\frac{\partial x^j}{\partial u_i}$, $x^j(0)=0$
and $\sum_{j=1}^n\frac{\partial x^j}{\partial u_i}(0)\frac{\partial x^j}{\partial u_k}(0)=\delta_{ik}$, such that $\psi_{ij}(u,\lambda)$ has the expansion
\[
\psi_{i1}(u,0)\psi_{ij}(u,\lambda)=\sum_{k=0}^\infty \frac{\partial\xi_k^j(u)}{\partial u_i}\lambda^k,\ \text{with } \xi_0^j(u)=x^j(u).
\]
\begin{definition}
\label{def1}
The matrix $\Xi'$ and row-vector $\Xi$ are defined as:
\[
\begin{aligned}
\Xi'(u,\lambda)&=\left(\psi_{i1}(u,0)\psi_{ij}(u,\lambda))\right)_{ij},\\[2mm]
\lambda\Xi(u,\lambda)&=\left(\sum_{i=1}^n\psi_{i1}(u,0)\psi_{ij}(u,\lambda))\right)_{j}.
\end{aligned}
\]
\end{definition}
{}{}From the equations (\ref{D10}) and (\ref{D11})
one deduces
\[
\frac{\partial \Xi_j(u,\lambda)}{\partial u_i}=
\Xi'(u,\lambda)_{ij}
\]
and  $\Xi$ satisfies
\begin{equation}
\label{f5.10}
\begin{aligned}
\frac{\partial^2 \Xi(u,\lambda)}{\partial u_i\partial u_j}&
=\Gamma_{ij}^i(u)\frac{\partial \Xi(u,\lambda)}{\partial u_i}+
\Gamma_{ji}^j(u)\frac{\partial \Xi(u,\lambda)}{\partial u_j},\qquad i\ne j\\
\frac{\partial^2 \Xi(u,\lambda)}{\partial (u_i)^2}&
=\sum_{j=1}^n
\Gamma_{ii}^j(u)\frac{\partial \Xi(u,\lambda)}{\partial u_j} +
\lambda\frac{\partial^2 \Xi(u,\lambda)}{\partial u_i\partial u_j},
\end{aligned}
\end{equation}
where the Christoffel symbols are given by (\ref{D6})
and hence that
\begin{equation}
\label{f5.11}
\frac{\partial^2 \Xi(u,\lambda)}{\partial x^k \partial x^\ell}=
\lambda\sum_{m=1}^n c_{k\ell }^m(u)\frac{\partial \Xi(u,\lambda)}{\partial x^m}.
\end{equation}
Thus $\Xi$ is the generating series for the flat sections of the connection
$\nabla_k=\frac{\partial}{\partial x^k}-\lambda c_{k\ell}^m$:
\[
\Xi_j= \delta_{j1}\lambda^{-1}+x^j(u)+\sum_{i=1}^\infty \xi^j_i\lambda^i.
\]
Moreover
\begin{equation}
\label{theta-prod}
\begin{aligned}
\Xi'(u,\lambda)\Xi'(u,-\lambda)^T&=\sum_{i=1}^n h_i(u)^2E_{ii},\\
\lambda \Xi(u,\lambda)\Xi'(u,-\lambda)^T&=(h_1^2(u)\ h_2^2(u)\ \ldots\ h_n^2(u)).
\end{aligned}
\end{equation}
Since
$\frac{\partial \Xi(u,\lambda)}{\partial x^i}$ is a linear combination of
$\frac{\partial \Xi(u,\lambda)}{\partial u_k}$'s,
\[
\lambda \Xi(u,\lambda)
\frac{\partial\Xi(u,-\lambda)^T }{\partial x^k}
\]
is independent of $\lambda$, which means that all coefficients, except the constant
coefficient, are zero. In particularly  the coefficient of $\lambda^2$ gives:
\[
\xi_1^m(u)=-\frac{\partial \xi_2^1(u)}{\partial x^m}+\sum_{i=1}^n
x^i(u)\frac{\partial \xi^i_1(u)}{\partial x^m}.
\]
The coefficient  of $\lambda$ of (\ref{f5.11}) leads to
\[
\frac{\partial^2 \xi_1^m(u)}{\partial u_k \partial u_\ell}=c_{k\ell
}^m(u),
\]
hence $\frac{\partial F(u)}{\partial x^m}=\xi_1^m(u)$ and
we obtain the Theorem of \cite{AKrV} (see also \cite{vdL1}).
\begin{theorem}
\label{theor Akm}
The function $F(u)=F(x(u))$ defined by
\[
F(u)=-\frac{1}{2}\xi_2^1(u)+\frac{1}{2}\sum_{i=1}^n
x^i(u)\xi_1^i(u)
\]
satisfies equation (\ref{D9}).
\end{theorem}

\section[The Sato Grassmannian and the Constrained KP]
{The Sato Grassmannian and the Constrained KP Hierarchy}
\label{Sat}
In this section we describe the Sato Grassmannian. Consider the spaces
\[
\begin{array}[h]{lcl}
H_- &=& \lambda^{-1} \mathbb{C}[[\lambda^{-1}]] = \{ \sum^{\infty}_{j=1} a_j
\lambda^{-j} | a_j \in \mathbb{C}\} \quad
\mbox{and}\\ H_+ &=& \mathbb{C}[\lambda] = \{ \sum^m_{i=0} b_i \lambda^i | b_i
\in \mathbb{C}\}.
\end{array}
\]
Hence $H= H_+ \oplus H_-$ is the quotient field $\mathbb{C}((\lambda))$ of
$\mathbb{C}[[\lambda^{-1}]]$. 
On the space $H$ we have a bilinear form, viz.  if
$f(\lambda)=\sum_j a_j \lambda^j$ and $g(\lambda)=\sum_j b_j \lambda^j$ are in $H$, then we define
\begin{equation}
\label{hel1}
(f(\lambda),g(\lambda))=Res_{\lambda}f(\lambda)g(\lambda)= \sum_j a_jb_{-j-1}.
\end{equation}
Let
$p_+ : H
\rightarrow H_+$ be the projection
\[
p_+ (\sum a_j \lambda^j) = \sum_{j\geq 0} a_j \lambda^j.
\]
Then the Sato Grassmannian $Gr (H)$ consists of all linear subspaces of $W\subset H$
that are of a size comparable
to $H_+$, i.e., 
\[
Gr (H) = \left\{ W\subset H \left|
\begin{array}[h]{l}
 p_+: W\to H_+ \;\;\mbox{has a finite}\\
\mbox{dimensional kernel and cokernel}
\end{array} \right. \right\}.
\]
The space $Gr (H)$ has a subdivision into different components:
\[
Gr^{(k)} (H) = \{ W \in Gr (H)|\ \dim
(\mbox{Coker}(p_+ |_W))  -\dim (\mbox{Ker} (p_+ |_W)) =k\}.
\]
Clearly, the subspace $\lambda^{k} H_+$ belongs to $Gr^{(k)} (H)$ and one
easily verifies that this also
holds for all subspaces in $Gr(H)$ that project bijectively onto
$\lambda^{k}H_+$, i.e. all $W$ belonging
to the ``big cell''.
For $W\in Gr(H)$, let $W^{\perp}$ be the orthocomplement of $W$ in
$H$ w.r.t. the bilinear form (\ref{hel1}). Then,  with the above given description,  $W^{\perp}$ also belongs
to $Gr(H)$ see \cite{HL}.

We write  $x=t_1$, $t=(t_1,t_2,t_3,\ldots)$ and  $\hat t=(t_2,t_3,t_4,\ldots)$. Consider now
a wave function $\psi_{BA}$ of the $KP$-hierarchy and
its dual $\psi_{BA}^*$ with 
\[
\begin{aligned}
\psi_{BA}(x,\hat t,\lambda)=&\{\sum_{j \leq 0}
a_j(x,\hat t) \lambda^j\}\lambda^le^{x\lambda+
\sum_{i>1} t_i \lambda^i}\;\text{and}\\
\psi_{BA}^*(x,\hat t,\lambda)=&\{\sum_{m \leq 0}
b_m(x,\hat t) \lambda^m\}\lambda^{-l}e^{-x\lambda - \sum_{i>1} t_i \lambda^i}.
\end{aligned}
\]
We assume from now on in this section that there exists an $\alpha$ of
the form 
\begin{equation}
\label{1.14}
\alpha(x,0) = x^N + \sum_{j>N}a_j x^j.
\end{equation}
such that for all $m\leq 0$ and all $j\leq 0$
\begin{equation}
\label{1.15}
\alpha(x,\hat t)a_j(x,\hat t) \in \mathbb{C}[[x,\hat t]] \;\;\text{and}
\;\;\alpha(x,\hat t)b_m(x,\hat t) \in
\mathbb{C}[[x,\hat t]].
\end{equation}
{\it Note that throughout this section $x^j$ stands for the $j$-th power of $x$
and not for the flat coordinate $x^j$ of Section \ref{Darb}.}

Such wave functions are called {\it regularizable}, see \cite{Sh},\cite{HL}. For regularizable
wave functions the
Laurent series in $x$ of $\psi_{BA}$ and $\psi_{BA}^*$ have the form
\[
\psi_{BA}(x,\hat t,\lambda)= \sum_{j\geq -N}
w_j(\hat t,\lambda)x^j,\;\;\text{where}\;\;w_j(\hat t,\lambda)=
\sum_{l=-\infty}^{N_1}v_l\lambda^l,\;\;\text{with}\;\; v_l \in
\mathbb{C}[[\hat t]],
\]
\[
\psi_{BA}^*(x,\hat t,\lambda)= \sum_{j\geq -N}
w_j^*(\hat t,\lambda)x^j,\;\;\text{where}\;\;w_j^*(\hat t,\lambda)=\sum_{l=-\infty}^{N_2}v^*_l\lambda^l,\
;\;\text{with}\;\; v^*_l \in
\mathbb{C}[[\hat t]].
\]
and moreover
\[
W=Span\{ w_j(0,\lambda),\;j\geq -N\} \;\;\text{and}\;\; W^*=Span\{
w_j^*(0,\lambda),\;j\geq -N\}
\]
belong to $Gr(H)$. It was Sato who realized that the space $W$ determines
$\psi_{BA}$, for
according to \cite{Sh} there holds
\begin{proposition}
\label{T2}
The map that associates to a regularizable wave function $\psi_{BA}$ of the
$KP$-hierarchy the span
of the coefficients in $\hat t=0$ of the Laurent series of $\psi_{BA}$ in $x$
is a bijection between
this class of wave functions and $Gr(H)$. The wave functions that
satisfy the conditions in
(\ref{1.15}) for $\alpha=1$ correspond to the big cell. 
\end{proposition}
For
each $W \in Gr(H)$ we denote the wave function corresponding to $W$ by $\psi_W$. The dual wave function of $\psi_W$, which we denote by
$\psi_W^*$ can be characterized as follows \cite{Sh},
\cite{HL}: 
\begin{proposition}
\label{C1}
Let $W$ and $\tilde{W}$ be two subpaces in $Gr(H)$. Then $\tilde{W}$ is the
space $W^*$
 corresponding to the dual wave function, if and only if
$\tilde{W}=W^{\perp}$ with
$W^{\perp}$ the orthocomplement of $W$ w.r.t. the bilinear form (\ref{hel1}) on $H$. Moreover
\[
\left(\psi_W(t,\lambda),\psi_W^*(s,\lambda)\right)=0.
\]
\end{proposition}
Let $W\in Gr^{(k)}(H)$ then 
\[
\psi_W(t,\lambda)=P_W(t,\partial_x)e^{\sum_{j=1}^\infty t_j\lambda^j},
\qquad
\psi_W^*(t,\lambda)=P_W^{*-1}(t,\partial_x)e^{-\sum_{j=1}^\infty t_j\lambda^j},
\]
where $P_W(t,\partial_x)$ is an $k^{th}$ order pseudo-differential operator.
The corresponding KP Lax operator ${\cal L}_W$ is equal to
\begin{equation}
\label{g0}
{\cal L}_W(t,\partial_x)=P_W(t,\partial_x)\partial_xP_W^{-1}(t,\partial_x).
\end{equation}
{}{}From now on we will use the notation $\psi_W$ and ${\cal L}_W$ instead of
$\psi_{BA}$ and ${\cal L}$ whenever we want to emphasize its dependence on a 
point $W$ of the Sato Grassmannian $Gr(H)$.

Eigenfunctions $\Phi$ and adjoint eigenfunctions $\Psi$ of the KP Lax operator (\ref{g0}), 
see (\ref{iso-pp}), can be expressed in wave and adjoint wave functions, viz. there 
exist functions $f,g\in H$ such that
\begin{equation}
\label{g1}
\Phi(t)=\left(\psi_W(t,\lambda),f(\lambda)\right), \qquad
\Psi(t)=\left(\psi_W^*(t,\lambda),g(\lambda)\right).
\end{equation}
Such (adjoint) eigenfunctions induce elementary B\"acklund--Darboux transformations \cite{HL}. 
Assume that we have the following data $W\in Gr^{(k)}(H)$, $W^\perp$, $\psi_W(t,\lambda)$ 
and $\psi_W^*(t,\lambda)$, then the (adjoint) eigenfunctions (\ref{g1}) 
induce new KP wave functions:
\begin{equation}
\label{g2}
\begin{aligned}
\psi_{W'}(t,\lambda)&=\left(\Phi(t)\partial_x\Phi(t)^{-1}\right)\psi_{W}(t,\lambda),&\quad
\psi_{W'}^*(t,\lambda)&=\left(\Phi(t)\partial_x\Phi(t)^{-1}\right)^{*-1}\psi_{W}^*(t,\lambda),\\
\psi_{{W''}}(t,\lambda)&=\left(-\Psi(t)\partial_x\Psi(t)^{-1}\right)^{*-1}\psi_{W}(t,\lambda),&\quad
\psi_{{W''}}^*(t,\lambda)&=\left(-\Psi(t)\partial_x\Psi(t)^{-1}\right)\psi_{W}^*(t,\lambda),
\end{aligned}
\end{equation}
where
\begin{equation}
\label{g3}
\begin{aligned}
W'&=\{ w\in W|(w(\lambda),f(\lambda))=0\}\in Gr^{(k+1)}(H),\quad
{W'}^\perp=W^\perp + \mathbb{C}f,\\
{W''}&=W+\mathbb{C}g\in Gr^{(k-1)}(H),\qquad {{W''}}^\perp =
\{ w\in W^\perp|(w(\lambda),g(\lambda))=0\}.
\end{aligned}
\end{equation}
Now assume that we have a Lax operator ${\cal L}_W$ with $W\in Gr^{(k)}(H)$ 
of the form (\ref{lax}), (\ref{iso-pp}), with $m$ minimal.
Then
\[
\begin{aligned}
\psi_{\lambda W}=&\lambda\psi_{W}\\
=&{\cal L}_W\psi_W\\
=&\left(\partial_x + \sum_{i=1}^{m} \Phi_i 
{\partial_x}^{-1} \Psi_i \right)\psi_{W}\\
=&\partial_x \left (\psi_{W}\right)+ \sum_{i=1}^{m} \Phi_i \Psi_i
\left(-\Psi_i\partial_x\Psi_i^{-1}\right)^{*-1} \psi_{W}\\
=&\partial_x \left (\psi_{W}\right)+ \sum_{i=1}^{m} \Phi_i \Psi_i\psi_{W_i},
\end{aligned}
\]
where $W\subset W_i$ of codimension 1. Hence
there exists a $W'=W+\sum_{i=1}^m W_i=W+\lambda W$ such that 
\begin{equation}
\label{g4}
\begin{aligned}
W\in  Gr^{(k)}(H),\qquad W'=W+\lambda W\in  Gr^{(k-m)}(H),\\
 W\subset W'\ 
(\text{codimension }m),\qquad \lambda W\subset W'\ 
(\text{codimension }m+1).
\end{aligned}
\end{equation}
The converse is also true. If ${\cal L}_W$ is a Lax operator such that $W$ satisfies (\ref{g4}), then there exists an inverse B\"acklund--Darboux transformation, i.e. the inverse of an $m^{th}$ order differential operator $L_m$, mapping $W$ into $W'$ and an $m+1^{th}$ order  B\"acklund--Darboux transformation $L_{m+1}$ mapping $W'$ into $\lambda W$:
\begin{equation}
\label{g5}
L_m^{-1}\psi_W=\psi_{W'},\qquad L_{m+1}\psi_{W'}=\psi_{\lambda W}=\lambda \psi_W.
\end{equation}
Hence,
\begin{equation}
\label{g6}
{\cal L}_W=L_{m+1}L_m^{-1},
\end{equation}
is a first order pseudo-differential operator. 
Before we continue, we first state a small Lemma which will be important
later on:
\begin{lemma}
\label{Lg1}
There exist $m$ independent functions $v_i\in W'$, $v_i\not\in W$, respectively 
$u_i\in W^\perp$, $u_i\not\in {W'}^\perp$ and  
 $m+1$ independent functions $f_j\in \lambda^{-1} W^\perp$, $f_j\not\in {W'}^\perp$, 
 respectively $g_j\in \lambda^{-1} W'$, $g_j\not\in W'$  
such that the functions 
$\varphi_i$ span $Ker (L_m)$, $\psi_i$ span $Ker (L_m^*)$,  
$\bar\varphi_j$ span $Ker(L_{m+1})$ and $\bar\psi_j$ span $Ker(L_{m+1}^*)$,  for $1\le i\le m$, $1\le j\le m+1$, where
\[
\begin{aligned}
\varphi_i(t)=&\left(\psi_{W'}(t,\lambda),u_i(\lambda)\right),\qquad
\psi_i(t)=\left(\psi_{W}^*(t,\lambda),v_i(\lambda)\right),\\[2mm]
\bar\varphi_j(t)=& \left(\psi_{W'}(t,\lambda),f_j(\lambda)\right),\qquad
\bar\psi_j(t)=\left(\psi_{W}^*(t,\lambda),g_j(\lambda)\right).
\end{aligned}
\] 
\end{lemma}
{\bf Proof} This follows from  (\ref{g2}--\ref{g4}) and 
the observation that $(\lambda W)^\perp=\lambda^{-1}W^\perp$.\hfill {$\square$}

\vskip 10pt

We proceed to construct ${\cal L}_W$.
Choose independent vectors $v_i\in W'=W+\lambda W$, as in Lemma \ref{Lg1},
such that  
$$
{W'}^\perp
=\{ w\in W^\perp | (w(\lambda),v_j(\lambda))=0\ \text{for all } 1\le j\le m \}.
$$
Since ${\cal L}_W$ can be obtained by two   B\"acklund--Darboux transformation $L_m$ and $L_{m+1}$ such that (\ref{g6}) holds, the general theory of such  B\"acklund--Darboux transformations \cite{HL} shows that
$$
({\cal L}_W)_-=\sum_{j=1}^m a_j(t)\partial_x^{-1}\left(\psi_W^*(t,\lambda),v_j(\lambda)\right).
$$
Thus
$$
\lambda \psi_W(t,\lambda)=\partial_x\left(\psi_W(t,\lambda)\right)+\sum_{j=1}^m a_j(\psi_W^*(t,\lambda),v_j)\psi_{W+\mathbb{C}v_j}(t,\lambda)
$$
Let now
$$
U_j=W+\mathbb{C}v_1+\mathbb{C}v_2+\cdots+\mathbb{C}v_{j-1}
     +\mathbb{C}v_{j+1}+\cdots+\mathbb{C}v_m .
$$
Choose $u_j\in W^\perp$ such that
$$
U_j=\{ w\in W'| (w(\lambda), u_j(\lambda))=0 \}.
$$
Then
$$
\begin{aligned}
(\lambda\psi_W(t,\lambda), u_i(\lambda))&=
\left(\lambda\psi_W(t,\lambda)-\partial_x\left(\psi_W(t,\lambda)\right), u_i(\lambda)\right)\\
=&\sum_{j=1}^m a_j(\psi_W^*(t,\lambda),v_j(\lambda))(\psi_{W+\mathbb{C}v_j}(t,\lambda),u_i(\lambda))\\
=&a_i(\psi_W^*(t,\lambda),v_i(\lambda))(\psi_{W+\mathbb{C}v_i}(t,\lambda),u_i(\lambda)),
\end{aligned}
$$
from which we can deduce that
$$
a_i=\frac{(\lambda\psi_W(t,\lambda), u_i(\lambda))}
{(\psi_W^*(t,\lambda),v_i(\lambda))(\psi_{W+\mathbb{C}v_i}(t,\lambda),u_i(\lambda))}.
$$
Since eigenfunctions which produce elementary B\"acklund--Darboux transformations are unique 
upto a scalar factor \cite{HL},\cite{HL2}, 
\[
\begin{aligned}
\psi_{W+\mathbb{C}v_i}(t,\lambda)&=\left((\psi^*_W(t,\lambda),v_i(\lambda))^{-1}\partial_x(\psi^*_W(t,\lambda),v_i(\lambda))\right)^{-1}\psi_{W}(t,\lambda),\\[2mm]
\psi_{W}(t,\lambda)&=\left( (\psi_{W+\mathbb{C}v_i}(t,\lambda),u_i(\lambda))\partial_x(\psi_{W+\mathbb{C}v_i}(t,\lambda),u_i(\lambda))^{-1}
\right) \psi_{W+\mathbb{C}v_i}(t,\lambda)
\end{aligned}
\]
and 
\[
\frac{\partial}{\partial t_n}(\psi^*_W(t,\lambda),v_i(\lambda))^{-1}=
\left({\cal L}^n_{W+\mathbb{C}v_i}\right)_+\left((\psi^*_W(t,\lambda),v_i(\lambda))^{-1}\right),
\]
we find that
$$
(\psi_W^*(t,\lambda),v_i(\lambda))^{-1}=c_i(\psi_{W+\mathbb{C}v_i}(t,\lambda),u_i(\lambda))
$$
and hence
$$
a_i=d_i(\lambda\psi_W(t,\lambda), u_i(\lambda)).
$$
Now replace 
$d_iu_i$ by $u_i$ and we obtain that
$$
\begin{aligned}
({\cal L}_W)_-&=\sum_{i=1}^m(\lambda\psi_W(t,\lambda), u_i(\lambda))\partial_x^{-1}(\psi_W^*(t,\lambda),v_i(\lambda))\\
=&\sum_{i=1}^m(\psi_W(t,\lambda), \lambda u_i(\lambda))\partial_x^{-1}(\psi_W^*(t,\lambda),v_i(\lambda)).
\end{aligned}
$$
With these choices, we set
\begin{equation}
\label{pp}
\Phi_i(t)=(\psi_W(t,\lambda), \lambda u_j(\lambda) ), \quad
\Psi_i(t)=(\psi_W^*(t,\lambda), v_j(\lambda) )
\end{equation}
and thus one obtains the desired Lax operator of the form given in 
(\ref{lax}).
%(\ref{lax}).
Notice that $(v_j(\lambda), u_i(\lambda))=0$ for $i\ne j$ and $(v_i(\lambda),u_i(\lambda))\ne 0$.
This construction for the Segal-Wilson Grassmannian was given in \cite{HL1}, see also \cite{HL2}.

For the Baker-Akhiezer wave function $\psi_{BA}$ the linear spectral problem 
${\cal L}  \psi_{BA} = \lambda \psi_{BA}$ 
can be decomposed on a set of differential equations :
\begin{equation}
\partial_x \psi_{BA}(t, \lambda)  + \sum_{i=1}^{m} \Phi_i \Gamma_i (t, \lambda) =  
\lambda \psi_{BA}(t, \lambda)  \;\; ; \;\;
\partial_x  \Gamma_i (t, \lambda) = \Psi_i (t) \psi_{BA} (t, \lambda) \, .
\label{two-BA}
\end{equation}
Similarly, we introduce the conjugated linear problem for
${\cal L}^{*}= - \partial_x - \sum_{i=1}^{m} \Psi_i
\partial_x^{-1}  \Phi_i$. For the conjugated Baker-Akhiezer wave function 
\begin{equation}
\psi_{BA}^{*} (t, \lambda) = \frac{ \tau (t+[\lambda^{-1}]) 
 }{ \tau (t)}  e^{- \sum_{j=1}^{\infty} \lambda^{j} t_j} \, ,
\label{ba-conj-fct}
\end{equation}
the conjugated spectral problem ${\cal L}^{*} (  \psi_{BA}^{*}) =
\lambda \psi_{BA}^{*}$ can be rewritten as :
\begin{equation}
\partial_x \Gamma_i^{*} (t,\lambda) = \Phi_i (t) \psi_{BA}^{*} (t,\lambda)
\;\; ; \; \; -\partial_x \psi_{BA}^{*}(t,\lambda) - \sum_{i=1}^{m} \Psi_i \Gamma_i^{*} (t,\lambda) 
=  \lambda \psi_{BA}^{*} (t,\lambda) \, .
\label{conjl}
\end{equation}
Recall from \cite{sep,HL}, that in the Sato formalism the squared eigenfunction
potentials $\Gamma_i (t,\lambda)$ and $\Gamma_i^{*}  (t,\lambda)$ with $i=1,{\ldots} ,m$
are given by:
\begin{eqnarray}
\Gamma_i  (t,\lambda) &=& \frac{1 }{ \lambda} \Psi_i (t-[\lambda^{-1}]) 
\frac{ \tau (t-[\lambda^{-1}])   }{ \tau (t)}
 e^{ \sum_{j=1}^{\infty} \lambda^{j} t_j} \label{sep} \\
\Gamma_i^{*} (t,\lambda) &=& -\frac{1 }{ \lambda} \Phi_i  (t+[\lambda^{-1}]) 
\frac{ \tau (t+[\lambda^{-1}])}{ \tau (t)}
e^{-\sum_{j=1}^{\infty} \lambda^{j} t_j} \, .
\end{eqnarray}

The Lax operator formalism possesses additional commuting symmetry flows \cite{EOR95}:
\begin{definition}
\label{definition:sato-add-flows}   
Mutually commuting additional symmetry flows of the integrable hierarchy
defined in terms of the pseudo-differential calculus  are defined as :
\begin{equation}
\partial_{k,n}\, {\cal L}  =  \lbrack {\cal M}^{(n)}_k \, , \, {\cal
L} \rbrack    \quad ; \quad {\cal M}^{(n)}_k = \sum_{l=0}^{n-1}  {\cal L}^l (\Phi_k)
\partial_x^{-1} {\cal L}^{*\,n-1-l} (\Psi_k)  \, .
\label{add-lax}
\end{equation}
for $n=1,2,{\ldots} $ and $k=1,{\ldots} ,m$.
\end{definition}
These flows induce :
\begin{eqnarray}
\partial_{k,n} {\cal L}^r (\Phi_i) &=& \sum_{l=0}^{n-1}  {\cal L}^l (\Phi_k) \partial_x^{-1}
\left( {\cal L}^{*\,n-1-l} (\Psi_k)  {\cal L}^r (\Phi_i) \right)
- {\cal L}^{r+n} (\Phi_i) \delta_{ki} \label{add-phi}\\
\partial_{k,n} {\cal L}^{*\,r} (\Psi_i) &=& \sum_{l=0}^{n-1}  {\cal L}^{*\,l}
(\Psi_k) \partial_x^{-1} \left( {\cal L}^{n-1-l} (\Phi_k)  {\cal L}^{*\,r}
(\Psi_i) \right) + {\cal L}^{*\,r+n} (\Psi_i) \delta_{ki} \, . \label{add-psi}
\end{eqnarray}
Especially, for $n=1$ we get :
\begin{equation}
\partial_{k,1} {\cal L}^r\left(\Phi_i\right) =  \Phi_{k} \beta_{ki}^{(r)}
- {\cal L}^{r+1} (\Phi_{k}) \delta_{ik}  \quad ; \quad
\partial_{k,1} {\cal L}^{*\,r} \left(\Psi_i\right) =  
\Psi_k \beta_{ik}^{*\, (r)}
+{\cal L}^{*\, r+1} (\Psi_{k})\delta_{ik} 
\label{eigenf-higher}
\end{equation}
where
\begin{equation}
\beta_{ij}^{(k)}\equiv \partial_x^{-1} \left( {\cal L}^{k} \,( \Phi_j) \Psi_i \right) \;\; ; \;\;
\beta_{ij}^{*\,(k)}\equiv \partial_x^{-1} \left(  \Phi_j \, {\cal L}^{*\,k} (
\Psi_i) \right)
\label{beta-def}
\end{equation}
with notation that for $k=0$ we write $\beta_{ij}=\beta_{ij}^{(0)}$.

Due to $ \sum_{k=1}^{m} {\cal M}_k^{(n)} = ({\cal L}^{n})_{-}$ we have
\begin{equation}
 \left( \frac{\partial }{ \partial t_n} +\sum_{k=1}^{m} \partial_{k,n}  \right) {\cal L} 
\, =\,\lbrack {\cal L}^n \, , \,  {\cal L}  \rbrack 
=0
\label{sum}
\end{equation}
or 
\begin{equation}
\sum_{k=1}^{m+1} \partial_{k,n}{\cal L} = 0 
\quad {\rm with} \quad  \partial_{{m+1},n} \equiv \frac{\partial }{ \partial
t_n} \, .
\label{sumi}
\end{equation}
Accordingly, it holds that :
\begin{equation}
\sum_{i=1}^{m+1} \partial_{i,n} \, \Phi_j = 0 \quad , \quad 
\sum_{i=1}^{m+1} \partial_{i,n}\, \Psi_j = 0 \;\; ; \;\; j=1, {\ldots} , m
\label{sumpp}
\end{equation}

We now extend the definition \ref{definition:sato-add-flows} 
to
\begin{equation}
\partial_{k,n} {\cal L}^{-1}  =  \lbrack {\cal M}^{(n)}_k \, , \, {\cal
L}^{-1} \rbrack  \, .
\label{add-lax-inv}
\end{equation}
Recall, from (\ref{g6}) and e.g. \cite{AGNP00,AFGZ00}, that 
$ {\cal L}^{-1}  = \sum_{j=1}^{m+1}  L_m ({\bar {\varphi}}_j) \partial_x^{-1}
{\bar \psi}_j$  in terms of 
$\left\{ {\bar {\varphi}}_j \right\}_{j=1}^{m+1}$ and
$\left\{ {\bar {\psi}}_j \right\}_{j=1}^{m+1}$ in $Ker (L_{m+1})$ and 
$Ker(L_{m+1}^\ast)$.

We introduce notation :
%$u_i \equiv u_i^{(n=1)}$ together with
\begin{equation}
 \Phi^{(-n)}_j = {\cal L}^{-n+1} \left( L_m ({\bar {\varphi}}_j) \right),
 \;\;\; \Psi^{(-n)}_j = {\cal L}^{*\, -n+1} \left( {\bar {\psi}}_j \right),
\quad j=1,{\ldots} ,m+1,\quad n=1,{\ldots} \, .
\label{phineg}
\end{equation}
Clearly, $ \Phi^{(-1)}_j= L_m ({\bar {\varphi}}_j)$,
${\cal L} (\Phi^{(-1)}_j)=0$ and ${\cal L}^{*} (\Psi^{(-1)}_j)=0$.

As a result of the definition (\ref{add-lax-inv}) (with $n=1$ ) we obtain :
\begin{equation}
\partial_{i,1} \Phi^{(-n)}_j = 
\Phi_i  \partial_x^{-1}  \left( \Psi_i \Phi^{(-n)}_j  \right) ,
\quad \partial_{i,1} \Psi^{(-n)}_j = 
-\Psi_i  \partial_x^{-1}  \left( \Phi_i \Psi^{(-n)}_j  \right) , 
\label{diopnnj}
\end{equation}
for $i=1,{\ldots} ,m,\quad j=1,{\ldots} , m+1$.
\begin{definition}
\label{definition:fsphi}
Define the $(m+1) \times (m+1)$ matrix $M =(M_{ij})_{1\le i,j\le m+1}$ by
\begin{equation}
M_{m+1\,j} = \sum_{n=1}^{\infty} \lambda^{n-1} \Phi^{(-n)}_{j}, \;\; \;   
M_{ij} = \partial_x^{-1}  \left( \Psi_i M_{m+1\,j}  \right),\quad
i=1,{\ldots} ,m ,\;\;  j=1,{\ldots} , m+1
\label{mdefs}
\end{equation}
\end{definition}
As pointed out in \cite{AFGZ00}, due to the fact 
that ${\cal L} (\Phi^{(-1)}_j)=0$, $M_{m+1\, j}$'s satisfy 
\begin{equation}
{\cal L} (M_{m+1\, j})  = \lambda  M_{m+1\, j} \,  
\label{laxfj}
\end{equation}
and in view of (\ref{diopnnj}) this can be rewritten as
\begin{equation}
\left( \partial_x + \sum_{i=1}^m  \partial_{i,1} \right) M_{m+1\,j}  = \lambda
M_{m+1\,j} \, .
\label{rew-laxfj}
\end{equation}
It follows that
\begin{equation}
  \partial_{i,1} M_{kj} =   \left\{ \begin{array}{ccc}
\partial_x^{-1}  \left( \Psi_k \Phi_j  \right) M_{ij}& {\rm for} & i\ne k=1,{\ldots} ,m\\
 \Phi_i \, M_{ij}   & {\rm for} & k=m+1, \,i=1,{\ldots} , m
\end{array} \right.
\label{paimkj}
\end{equation}
for arbitrary $j=1,{\ldots} , m+1$.
Furthermore, due to (\ref{mdefs}) and (\ref{rew-laxfj}) we find :
\begin{equation}
\label{mijlamb}
\begin{aligned}
\partial_x M_{ij} = \Psi_i M_{m+1 \,j}, \quad i=1,{\ldots} ,m ,\\
\left( \partial_x - \lambda \right) M_{m+1 \,j}
 + \sum_{i=1}^m  \Phi_{i}  M_{ij}  = 0,
\;\;\; j = 1, {\ldots}, m+1 \, .
\end{aligned}
\end{equation}
As will be shown in the next section, this matrix will appear in the
Riemann-Hilbert problem and together with the dressing matrix will turn out
to be a crucial element in the integrable structure 
behind the Darboux-Egoroff system.

\section[Dressing Formalism and Integrable Hierarchy]
{Dressing Formalism and Integrable Hierarchy in the 
  Homogeneous Gradation}
\label{DF}
In this section we first briefly describe the Riemann-Hilbert problem
(see, for instance \cite{AKJ01} and references therein) 
resulting in the integrable hierarchy associated to the affine Lie algebra
${\cal G} = {\widehat {sl}} (m+1)$ with the homogeneous gradation.
We then derive an expression for the dressing matrix (see references
\cite{AGNPZ} and references therein) in terms of
underlying $\tau$- and potential-functions and additional symmetry
flows by comparing with the results obtained for the same hierarchy
in the previous section.

Let $G$ be a Lie group associated to the Lie algebra
${\cal G} = {\widehat {sl}} (m+1)$ with the homogeneous gradation. 
We define two subgroups of $G$ as :
\begin{equation}
G_{-} = \{g \in G | g(\lambda) = 1+\sum_{i<0} g^{(i)}
\}\, ,
\label{gmin}
\end{equation}
\begin{equation}
G_{+} = \{g \in G | g(\lambda) = \sum_{i\geq 0} g^{(i)}
\}\, ,
\label{gplus}
\end{equation}
where $g^{(i)}$ has gradation $i$ with respect to the gradation operator
$d= \lambda d /d \lambda$. Also $G_{+}\cap G_{-}=I$.

\begin{definition}
\label{definition:rh-problem}    
The (extended) Riemann-Hilbert problem for ${\cal G} = {\widehat {sl}} (m+1)$
with the homogeneous gradation is defined as \cite{AGZ01}:
\begin{equation}
\exp \left({\sum_{j=1}^{m+1} \sum_{n=1}^{\infty} E^{(n)}_{jj}u^{(n)}_j} \right)g
= g_{-}g_{+} = \Theta^{-1} M
\label{rh-def}
\end{equation}
with $g$ being a constant element in $G_{-}G_{+}$,
$( E_{rs})_{ij} = \delta_{ir} \delta_{js}$ and
$E^{(n)}_{jj}= \lambda^n E_{jj}$.
\end{definition}
We find:
\begin{equation}
\frac{\partial}{\partial u^{(n)}_j}
\, \exp \left({\sum_{j=1}^{m+1} \sum_{n=1}^{\infty} E^{(n)}_{jj}u^{(n)}_j}\right) g
= \left( \frac{\partial}{\partial u^{(n)}_j} g_{-} (u) \right) g_{+} (u) 
+ g_{-} (u) \frac{\partial}{\partial u^{(n)}_j} g_{+} (u)
\label{rh-tk}
\end{equation}
or
\begin{equation} 
 g_{-}^{-1} (u)  E^{(n)}_{jj} g_{-} (u) = g_{-}^{-1}(u) 
\left( \frac{\partial}{\partial u^{(n)}_j} g_{-} (u) \right)
+\left( \frac{\partial}{\partial u^{(n)}_j} g_{+} (u) \right)
 g_{+}^{-1}(u) \, .
\label{rh-tka}
\end{equation}
Note, that $g_{-}^{-1}
\left( \frac{\partial}{\partial u^{(n)}_j} g_{-} \right)$ is in 
${\cal G}_{-}$
and $\left( \frac{\partial}{\partial u^{(n)}_j} g_{+}  \right)
 g_{+}^{-1}$ is in ${\cal G}_{+}$
where ${\cal G}_{\pm }$ are positive/negative subalgebras of the graded 
Lie algebra ${\cal G}$.
The identity (\ref{rh-tka}) implies therefore that:
\begin{equation}
\frac{\partial}{\partial u^{(n)}_j}\, g_{-} = g_{-} 
\left(g_{-}^{-1}E^{(n)}_{jj} g_{-} \right)_{-} \, .
\label{ugmin}
\end{equation}
Let $({\bf u})= ({\bf u_1}, {\ldots} , {\bf u_{m+1}})$ denote $m+1$ multi-times
(or flows) ${\bf u_j}$ with each argument ${\bf u_j}$ denoting one of the $m+1$ multi-times
$(u_j^{(1)}, u_j^{(2)}, {\ldots} )$.
%and where ${\bf u_{m+1}}$ denotes for convenience 
%the ``usual'' isospectral multi-time $t$.

In terms of the $\Theta ({\bf u}, \lambda) =g_{-}^{-1}({\bf u}, \lambda) $ we
obtain the following lemma :
\begin{lemma}
\label{lemma:add-flows}
Mutually commuting $u^{(n)}_j$-flows in the dressing formalism
of the integrable hierarchy are defined through :
\begin{equation}
\frac{\partial  }{ \partial  u^{(n)}_j} \, \Theta ({\bf u}, \lambda) 
= - (\Theta \lambda^n E_{jj}{\Theta}^{-1})_{-} \Theta ({\bf u}, \lambda) 
\quad ; \quad j =1, {\ldots} ,m+1 \, .
\label{snj-def}
\end{equation}
\end{lemma}
Let us note, that inserting the non-diagonal matrices $E_{ij}$ with $i \ne j$
on the right hand side of (\ref{snj-def}) would correspond to nonabelian
symmetry flows from the Borel loop subalgebra of ${\widehat {sl}} (m+1)$,
described recently in \cite{AGNPZ,AGNP00}.

For $M ({\bf u}, \lambda) =g_{+} ({\bf u}, \lambda) $ we obtain
from equation (\ref{rh-tka}) :
\begin{equation}
\frac{\partial}{\partial u^{(n)}_j} M ({\bf u}, \lambda) = 
\left(\Theta E^{(n)}_{jj} \Theta^{-1} \right)_{+} M ({\bf u}, \lambda) 
\label{umplus}
\end{equation}

Consider (\ref{umplus}) for $n=1$ and $j=m+1$. {}From now on we will identify
$u^{(1)}_j=u_j$.
The result can be written (see also \cite{vdL1}) as :
\begin{equation}
\left(\frac{\partial}{\partial u_{m+1}} - \lambda E_{m+1\, m+1} +
\lbrack \theta^{(-1)}, E \rbrack \right) M =0 \, , 
\label{umnone}
\end{equation}
where  $E$ is a semisimple, grade-one
element of ${\cal G}$ 
\begin{equation}
E \equiv \frac{\lambda  }{ m+1} I - \lambda E_{m+1 \, m+1}
\label{defE}
\end{equation}
and 
$\theta^{(-1)}$ is a term of $\Theta=1 + \theta^{(-1)}+ \ldots$ of grade $-1$.
Since the grade-zero matrix $A \equiv \lbrack \theta^{(-1)}, E \rbrack $ 
is in the image of $ad (E)$ it can therefore be parametrized as :
\begin{equation}
A  = \sum_{i=1}^{m} \left( - \Psi_i \, 
E_{i \, m+1} + \Phi_i \, E_{m+1 \, i}\right)
\, .
\label{a-homo}
\end{equation}

A special role in this formalism is played by the $u^{(n)}_{m+1}$-flows. 
{}{}From (\ref{defE})
we find that :
\begin{equation}
\frac{\partial  }{ \partial  u^{(n)}_{m+1}}\, \Theta  ({\bf u}, \lambda) 
=  (\Theta \lambda^{n-1} E {\Theta}^{-1})_{-} \Theta  ({\bf u}, \lambda) \, ,
\label{snmp1-def}
\end{equation}
which shows that the $u^{(n)}_{m+1}$-flows are the isospectral deformations
of the underlying integrable hierarchy \cite{AGNPZ}.
Consequently,
\begin{equation}
u^{(n)}_{m+1}=t_n \quad {\rm and}\quad x=t_1= u_{m+1} \,.
\label{iso=u}
\end{equation}
The standard solution to equation (\ref{umnone}) was given in the literature
in form of the formal path ordered integral (see e.g. \cite{AFGZ00} and
references therein).
Here we notice, that with the above identification and with parametrization as in 
equation (\ref{a-homo}) relation (\ref{umnone}) agrees with equation (\ref{mijlamb}) 
for the matrix $M$ introduced in the previous section.

An important property :
\begin{equation}
\sum_{j=1}^{m+1} \frac{\partial  }{ \partial  u^{(n)}_j} \, \Theta = 0 \, ,
\quad n=1, {\ldots} 
\label{snj-sum}
\end{equation}
is another consequence of lemma \ref{lemma:add-flows}.

For $\Theta$ we obtain the dressing expression:
\begin{equation}
\Theta^{-1}  \left(\frac{\partial}{\partial u_{m+1}} +E
%+ \lbrack \theta^{-1}, E \rbrack 
+ A
\right) \Theta = \frac{\partial}{\partial u_{m+1}} +E \, , 
\label{dressa}
\end{equation}
by taking $n=1$ in equation (\ref{snj-def}).
Hence we have shown that $Ad (\Theta)$ maps the matrix operator
$ \partial_x + E$ with $\partial_x = \frac{\partial}{\partial u_{m+1}}$
into the Lax operator :
\begin{equation}
    L= \partial_x + E +A \, , 
\label{lax-matrix}    
\end{equation}
according to 
\begin{equation}
\partial_x + E    \to \Theta (  \partial_x + E )
\Theta^{-1} =  \partial_x + E +A \, .
\label{dress-def}
\end{equation}
This is called the dressing procedure \cite{AGNPZ}.

One also finds that the action of the $u^{(n)}_{j}$-flows on the potential
$A$ is given by :
\begin{equation}
\frac{\partial  }{ \partial  u^{(n)}_j} \, A = \lbrack  (\Theta \lambda^n E_{jj}
{\Theta}^{-1})_{+}  \, , \, L \rbrack \, .
\label{symm-ona}
\end{equation}
The $u^{(n)}_{j}$-flows are the symmetry flows of the underlying integrable
hierarchy due to the fact that they commute with the isospectral flows.
As explained in \cite{AGNPZ}, the dressing formalism is connected with the
tau-function $\tau$ being a function of all the $u^{(n)}_{j}$-times.
The connection is expressed by the formula :
\begin{equation}
{\rm   Res}_{\lambda} \left( {\rm tr} ( E \Theta \lambda^{n-1} E_{jj}
\Theta^{-1} ) \right)\, =\,
- \frac{\partial  }{ \partial  u^{(n)}_j} \partial_x \ln \tau
({\bf u}) \, .
\label{papatau}
\end{equation}
%Also, ${\rm   Res}_{\lambda} ( {\ldots} )$ is a projection on the coefficient
%of $\lambda^{-1}$ in $( {\ldots} )$.

Our goal in this section is to derive expression for the dressing matrix $\Theta$ in terms
of the $\tau({\bf u})$ function and the matrix elements
$\Phi_i, \Psi_i$ of $A$ from (\ref{a-homo}).
A simple way to do it relies on equivalence between the
above algebraic formulation and the one in which the above integrable model 
is represented in the Sato formalism by the pseudo-differential Lax operator
as in the previous section. In reference \cite{AGZ95} this equivalence was
established by showing that recursion operators of both hierarchies are 
identical.
A key to our derivation is the following Proposition, which shows that the
additional symmetry flows of both approaches agree.

\begin{proposition}
\label{proposition:identify}
We have    
\begin{equation}
\frac{\partial }{ \partial u^{(n)}_i} = \partial_{i,n} 
\quad i=1, {\ldots} ,m+1 \, .
\label{agree}
\end{equation}
Meaning, that both flows have identical actions on the fields
$\Phi_i, \Psi_i,\; i=1,{\ldots}, m$ parametrizing the constrained KP
hierarchy.
\end{proposition}
{\bf Proof} For $i=m+1$ this was already established in relation (\ref{iso=u}).
For other values of index $i$ the relation follows from the direct
calculation based on (\ref{symm-ona}).\hfill {$\square$}

\vskip 10pt

We will now use the equivalence of these two formalisms to calculate
the dressing matrix $\Theta$. For simplicity, we first will work with
$m=2$ for which case the  Lax matrix operator in (\ref{lax-matrix})
becomes :
\begin{equation}
L = \partial_x  + E + A = \partial_x + 
\frac{\lambda }{ 3}\threemat{1}{0}{0}{0}{1}{0}{0}{0}{-2} +
\threemat{0}{0}{-\Psi_1}{0}{0}{-\Psi_2}{\Phi_1}{\Phi_2}{0}\, .
\label{lax-sl3}
\end{equation}
We start with the corresponding un-dressed linear spectral problem:
\begin{equation}
( \partial_x  + E ) \exp (-\sum E^{(n)} t_n) \,  \chi = 0 \, , 
\label{linear}
\end{equation}
where $\chi$ is any constant column vector, $E=E^{(1)}$
and
\begin{equation} 
E^{(n)} \equiv \frac{\lambda^n }{ 3}\threemat{1}{0}{0}{0}{1}{0}{0}{0}{-2} \, .
\label{e-matrix}
\end{equation}
We will be working with three types of $\chi$ :
\begin{equation}
\chi_1 = \left(\begin{array}{c}  
1\\ 0\\ 0 \end{array} \right) \;\; ; \;\; 
\chi_2 = \left(\begin{array}{c}  
0\\ 1\\ 0 \end{array} \right) \;\; ; \;\; 
\chi_3 = \left(\begin{array}{c}  
0\\ 0\\ 1 \end{array} \right) \; .
\label{chis}
\end{equation}
Choosing $\chi_3$ in (\ref{linear}) and multiplying from the left by $\Theta$
we get an information about the last (3-rd) column of the $\Theta$ matrix :
\begin{equation}
( \partial_x + E +A) \left(\begin{array}{c} \theta_{13}\\ \theta_{23}\\ 
\theta_{33} \end{array} \right)\exp \left((2/3)\sum \lambda^{n} t_n\right)
=0 \, .
\label{lin-th3}
\end{equation}
This equation takes the form of the linear spectral problem as in 
(\ref{two-BA}) with 
\begin{equation}
\theta_{13} = \Gamma_1 e^{-\sum \lambda^n t_n}  \;\; ; \;\;
\theta_{23} = \Gamma_2 e^{-\sum \lambda^n t_n}  \;\; ; \;\;
\theta_{33} =  \psi_{BA} e^{-\sum \lambda^n t_n} \, ,
\label{lastcol}
\end{equation}
for  $\Gamma_i, \, i=1,2$ as in equation (\ref{two-BA}).
For the choice $\chi=\chi_2$ in (\ref{linear}) we find :
\begin{equation}
\left( \partial_x + E +A\right) \left(\begin{array}{c} \theta_{12}\\ \theta_{22}\\ 
\theta_{32} \end{array} \right)\exp \left(-\sum \lambda^{n} t_n/3\right) =0
\, .
\label{lin-th}
\end{equation}
After multiplying from the left by $\left(-2 \sum \lambda^{n} t_n/3\right)$
we obtain :
\begin{equation}
\left( \partial_x + 
\threemat{\lambda}{0}{0}{0}{\lambda}{0}{0}{0}{0} +
\threemat{0}{0}{-\Psi_1}{0}{0}{-\Psi_2}{\Phi_1}{\Phi_2}{0} \right)
\left(\begin{array}{c} \theta_{12}\\ \theta_{22}\\ 
\theta_{32} \end{array} \right)\exp \left(-\sum \lambda^{n} t_n\right) =0 \, .
\label{linea-fin}
\end{equation}
In components this is equivalent to
\begin{eqnarray}
0&=&\partial_x \theta_{12} - \Psi_1 \theta_{32}  \label{paxthot}\\
0&=&\partial_x \theta_{22} - \Psi_2 \theta_{32}  \label{paxthtt}\\
\lambda \theta_{32} &=&\partial_x \theta_{32} + \Phi_1 \theta_{12}
+ \Phi_2 \theta_{22} \,.
\label{paxthtth}
\end{eqnarray}
Solving the first two equations we get :
\begin{eqnarray}
\theta_{12} &=& \partial^{-1}_x \left( \Psi_1 \theta_{32} \right) \label{paithtth}\\
\theta_{22} &=& 1+\partial^{-1}_x \left( \Psi_2 \theta_{32} \right) \, .
\label{paithttw}
\end{eqnarray}
Note, that the integration constant in (\ref{paithttw})
have been chosen so that the diagonal element
$\theta_{22}$ starts with $1$.

Plugging this back into (\ref{paxthtth}) we get an expression :
\begin{equation}
{\cal L} (\theta_{32}) = \lambda \theta_{32} - \Phi_2 
\label{thtwth}
\end{equation}
whose solution is :
\begin{equation}
\theta_{32} = \sum^{\infty}_{i=1} \lambda^{-i} {\cal L}^{i-1} ( \Phi_2) \, .
\label{thtwth-def}
\end{equation}
Correspondingly, 
\begin{equation}
\theta_{12} = \sum^{\infty}_{i=1} \lambda^{-i} \beta_{12}^{(i-1)} 
\;\; ; \; \; %\label{betto}\\
\theta_{22} = 1+\sum^{\infty}_{i=1} \lambda^{-i} \beta_{22}^{(i-1)}\, .\label{betttw}
\end{equation}
with $\beta_{ij}^{(k)}$ as in (\ref{beta-def}).

Similar technique yields for the choice $\chi=\chi_1$ :
\begin{equation}
\theta_{31} = \sum^{\infty}_{i=1} \lambda^{-i} {\cal L}^{i-1} ( \Phi_1)
\label{thoth-def}
\end{equation}
and 
\begin{equation}
\theta_{11} = 1+ \sum^{\infty}_{i=1} \lambda^{-i} \beta_{11}^{(i-1)} 
\;\; ; \; \; %\label{betoo}\\
\theta_{21} = \sum^{\infty}_{i=1} \lambda^{-i} \beta_{21}^{(i-1)} \, .
\label{betotw}
\end{equation}

These results complete the construction of the dressing matrix :
\begin{equation}
\Theta =
\threemat{1+\sum^{\infty}_{i=1} \lambda^{-i} \beta_{11}^{(i-1)}}
{\sum^{\infty}_{i=1} \lambda^{-i} \beta_{12}^{(i-1)}}
{\Psi_1 (t-[\lambda^{-1}])\frac{\tau (t-[\lambda^{-1}]) }{ \lambda \tau (t)}}
{\sum^{\infty}_{i=1} \lambda^{-i} \beta_{21}^{(i-1)}}
{1+\sum^{\infty}_{i=1} \lambda^{-i} \beta_{22}^{(i-1)}}
{\Psi_2 (t-[\lambda^{-1}])\frac{\tau (t-[\lambda^{-1}]) }{ \lambda \tau (t)}}
{\sum^{\infty}_{i=1} \lambda^{-i} {\cal L}^{i-1}(\Phi_1)}
{\sum^{\infty}_{i=1} \lambda^{-i} {\cal L}^{i-1}(\Phi_2)}
{\frac{\tau (t-[\lambda^{-1}]) }{ \tau (t)}}
\label{theta-m2}
\end{equation}
{}For the first two diagonal elements we find from \cite{ANP98,ANP99} :
\begin{equation}
1+\sum^{\infty}_{i=1} \lambda^{-i} \beta_{jj}^{(i-1)} = 
\frac{\tau ({\bf u_j}-[\lambda^{-1}]) }{  \tau (t)} \quad ;\quad j=1,2\, .
\label{diag-id}
\end{equation}
Note, that on the right hand side we have only showed 
the multi-times ${\bf u_j}$ which are shifted
according to $({\bf u_j}-[\lambda^{-1}])= (u_j^{(1)}-1/\lambda,
u_j^{(2)}-1/2\lambda^2,{\ldots} )$.

Moreover, from \cite{ANP97,ANP98,ANP99} we find for the first two elements of
the last row of $\Theta$:
\begin{equation}
\sum^{\infty}_{i=1} \lambda^{-i} {\cal L}^{i-1}(\Phi_j) = 
\Phi_j ({\bf u_j}-[\lambda^{-1}])\frac{\tau ({\bf u_j}-[\lambda^{-1}]) }{ \lambda \tau (t)}
\quad ;\quad j=1,2
\label{22-p}
\end{equation}
Relations in (\ref{22-p}) follow from the ones in (\ref{diag-id}) by
the Darboux-B\"{a}cklund transformation generated by 
$T_j =\Phi_j \partial_x  \Phi_j$ applied to the Lax operator :
\begin{equation}
{\tilde {\cal L}}  = T_j {\cal L} T_j^{-1} = \partial_x +\sum_{i=1}^2
{\tilde \Phi}_i \, .
\partial_x^{-1} {\tilde \Psi}_i
\label{l-ti}
\end{equation}
Inserting the transformed quantities into identity (\ref{diag-id})
one obtains relations (\ref{22-p}).

Another set of identities can be obtained from (\ref{diag-id})
using the binary Darboux-B\"{a}cklund transformations \cite{sep}. 
As an illustration we consider :
\begin{eqnarray}
{\cal L} &\to& {\tilde {\cal L}} = T_2 {\cal L} T_2^{-1}  \quad ;\quad T_2 = \Phi_2 \partial_x \Phi_2^{-1}
\label{abdbone}\\
{\tilde {\cal L}}  &\to& {\bar {\cal L}} = {\bar T}^{*\,-1}_1{\tilde {\cal L}} {\bar T}^{*}_1
= \partial_x +\sum_i {\bar \Phi}_i \partial_x^{-1} {\bar \Psi}_i  \;\; ;\;\;
{\bar T}_1 = {\tilde \Psi}_1 \partial_x {\tilde \Psi}_1^{-1}
\label{abdbtwo}
\end{eqnarray}
where 
\begin{eqnarray}
{\bar \Phi}_1 &=& \Phi_2 / \partial_x^{-1} (\Phi_2, \Psi_1) \;\; ;\;\;
{\bar \Psi}_1 = {\bar T}_1 { T}^{*\,-1}_2 {\cal L}^* (\Psi_1)
\label{bara-one}\\
{\bar \Phi}_2 &=& {\bar T}^{*\,-1}_1 T_2 {\cal L} (\Phi_2)
\;\; ;\;\;
{\bar \Psi}_2 = {\bar T}_1  (1 /\Phi_2)
\label{bara-two}
\end{eqnarray}
and $ {\tilde \Psi}_1= - \partial_x^{-1} (\Phi_2, \Psi_1) / \Phi_2$.
The $\tau$-function transforms as $\tau \mapsto - \partial_x^{-1} (\Phi_2\,
\Psi_1) \tau$
and for $j=2$ (\ref{diag-id}) becomes:
\begin{equation}
1+\sum^{\infty}_{i=1} \lambda^{-i} \partial_x^{-1} \left( {\bar {\cal
L}}^{i-1} ({\bar \Phi}_2)\, {\bar \Psi}_2 \right)  = 
\frac{\partial_x^{-1} (\Phi_2\, \Psi_1) ({\bf u_2} - [\lambda^{-1}])
  \tau ({\bf u_2}-[\lambda^{-1}]) }{
\partial_x^{-1} (\Phi_2\, \Psi_1) (t)  \tau (t)} \, .
\label{diag-id-bar}
\end{equation}
By inserting definitions (\ref{bara-one})-(\ref{bara-two}) one can now show that 
\begin{equation}
\partial_x^{-1} \left( {\bar {\cal L}}^{i-1} ({\bar \Phi}_2)\, {\bar \Psi}_2 \right)
= \frac{ \partial_x^{-1} \left( {\cal L}^i ({ \Phi}_2)\, { \Psi}_1 \right) }{ \partial_x^{-1} (\Phi_2, \Psi_1)}
\label{tech-ss}
\end{equation}
and the identity :
\begin{equation}
 \sum^{\infty}_{i=1} \lambda^{-i} \beta_{12}^{(i-1)} = 
\beta_{12} ({\bf u_2}-[\lambda^{-1}])\frac{\tau ({\bf u_2}-[\lambda^{-1}]) }{ \lambda \tau (t)} 
\label{susp-a}
\end{equation}
follows from (\ref{diag-id-bar})
after dividing both sides by $\lambda$ and multiplying by $\partial_x^{-1}
(\Phi_2\, \Psi_1)$.
Similarly, we obtain
\begin{equation}
 \sum^{\infty}_{i=1} \lambda^{-i} \beta_{21}^{(i-1)} = 
\beta_{21} ({\bf u_1}-[\lambda^{-1}])\frac{\tau ({\bf u_1}-[\lambda^{-1}])
}{ \lambda \tau (t)} \, .
\label{susp-b}
\end{equation}
To summarize, we have found the following dressing matrix:
\begin{equation}
\Theta = \frac{1 }{ \tau (t) }
\threemat{\tau ({\bf u_1}-[\lambda^{-1}])}
{\beta_{12} ({\bf u_2}-[\lambda^{-1}]) \frac{\tau ({\bf u_2}-[\lambda^{-1}])}{ \lambda}}
{\Psi_1 (t-[\lambda^{-1}])\frac{\tau (t-[\lambda^{-1}]) }{ \lambda }}
{\beta_{21} ({\bf u_1}-[\lambda^{-1}]) \frac{\tau ({\bf u_1}-[\lambda^{-1}])}{ \lambda}}
{\tau ({\bf u_2}-[\lambda^{-1}])}
{\Psi_2 (t-[\lambda^{-1}])\frac{\tau (t-[\lambda^{-1}]) }{ \lambda }}
{\Phi_{1} ({\bf u_1}-[\lambda^{-1}]) \frac{\tau ({\bf u_1}-[\lambda^{-1}])}{ \lambda}}
{\Phi_{2} ({\bf u_2}-[\lambda^{-1}]) \frac{\tau ({\bf u_2}-[\lambda^{-1}])}{ \lambda}}
{\tau (t-[\lambda^{-1}])} 
\label{theta-flows}
\end{equation}
% The inverse matrix is obtained along the same lines
% and equal to :
% \begin{equation} 
% \Theta^{-1} = 
% \threemat{\frac{\tau ({\bf u_1}+ [\lambda^{-1}]) }{ \tau (t) }}
% {-\beta_{12} ({\bf u_1}+[\lambda^{-1}]) \frac{\tau ({\bf u_1}+[\lambda^{-1}])}{ \lambda \tau (t)}}
% {-\Psi_{1} ({\bf u_1}+[\lambda^{-1}]) \frac{\tau ({\bf u_1}+[\lambda^{-1}])}{ \lambda\tau (t)}}
% {-\beta_{21} ({\bf u_2}+[\lambda^{-1}]) \frac{\tau ({\bf u_2}+[\lambda^{-1}])}{ \lambda\tau (t)}}
% {\frac{\tau ({\bf u_2}+ [\lambda^{-1}]) }{ \tau (t) }}
% {-\Psi_2 ({\bf u_2}+[\lambda^{-1}])\frac{\tau ({\bf u_2}+[\lambda^{-1}]) }{ \lambda\tau (t) }}
% {-\Phi_{1} (t+[\lambda^{-1}]) \frac{\tau (t+[\lambda^{-1}])}{ \lambda \tau (t)}}
% {-\Phi_{2} (t+[\lambda^{-1}]) \frac{\tau (t+[\lambda^{-1}])}{ \lambda \tau (t)}}
% {\frac{\tau (t+[\lambda^{-1}]) }{  \tau (t)}}
% \label{th-inv-flows}
% \end{equation}

We will now generalize the above matrix to the general case of $m \geq 2$.
For this purpose we introduce the following definition which extends
definition of coefficients $\beta_{ij}=\beta_{ij}^{(k=0)}$ previously given in 
(\ref{beta-def}) :
\begin{definition}
\label{definition:betas}    
Let
\begin{equation}
\beta_{ij} ({\bf u}) = \partial_x^{-1} \left(\Phi_j  \Psi_i \right) ({\bf u}) \;\; ;\;\;
\beta_{j \,m+1 } ({\bf u}) = \Psi_j ({\bf u}) \;\; ;\;\;
\beta_{m+1\, i} ({\bf u}) = \Phi_i ({\bf u}) \;\; 
\label{bets}
\end{equation}
for $i \ne j$ and $i,j=1, {\ldots} , m$.
\end{definition}
First, note that with this identification equations (\ref{paimkj}) and
(\ref{mijlamb}) take formally form of the Darboux-Egoroff system in (\ref{D10})
provided we also make coefficients $\beta_{ij}$ symmetric (see next section).

Moreover, based on the Definition \ref{definition:betas}  
we can now formulate the following proposition:
\begin{proposition}
\label{proposition:theta-matrix}
The matrix elements 
of the dressing matrix  $\Theta=(\theta_{ij})_{1\le i,j\le m+1}$ are given by :
\begin{eqnarray}
\theta_{k\ell} ({\bf u}, \lambda) & = & \beta_{k\ell} ({\bf u_\ell}-[\lambda^{-1}]) \frac{\tau
({\bf u_\ell}-[\lambda^{-1}])}{ \lambda \tau ({\bf u})} 
\quad \ell \ne k \;\;\; k,\ell =1, {\ldots} , m+1
\label{th-gena}\\
\theta_{kk} ({\bf u}, \lambda)  & = &  \frac{\tau ({\bf u_k}-[\lambda^{-1}])}{  \tau
({\bf u})}  \qquad 
k =1, {\ldots} , m+1 \label{th-genb}
\end{eqnarray}
while the matrix elements of the inverse dressing matrix 
$\Theta^{-1}=(\theta_{ij}^{-1})_{1\le i,j\le m+1}$ are :
\begin{eqnarray}
\theta_{k\ell}^{-1} ({\bf u}, \lambda) & = & -\beta_{k\ell} ({\bf u_k}+[\lambda^{-1}]) \frac{\tau
({\bf u_k}+[\lambda^{-1}])}{ \lambda \tau ({\bf u})} 
\quad \ell \ne k \;\;\; k,\ell =1, {\ldots} , m+1
\label{th-inv-gena}\\
\theta_{kk}^{-1} ({\bf u}, \lambda) & = &  \frac{\tau ({\bf u_k}+[\lambda^{-1}])}{  
  \tau ({\bf u})} \qquad 
k =1, {\ldots} , m+1 \label{th-inv-genb}
\end{eqnarray}
\end{proposition}

\begin{definition}
\label{definition:gamma-matrix}
The $\Gamma$ matrix is defined as :
\begin{equation}
\Gamma ({\bf u}, \lambda) = \Theta ({\bf u}, \lambda)\, D ({\bf u}, \lambda)  
\quad ; \quad  D_{ij} ({\bf u}, \lambda)  = \delta_{ij}  
\, e^{\sum^{\infty}_{n=1} \lambda^n u_j^{(n)}} 
\label{gms}
\end{equation}
or in components for $\Gamma=(\gamma_{ij})_{1\le i,j\le m+1}$
\begin{equation}
\gamma _{i\,k} ({\bf u}, \lambda) = \theta_{i\,k} ({\bf u}, \lambda) 
\; e^{\sum^{\infty}_{n=1} \lambda^n u_k^{(n)}}
\qquad k,i=1,{\ldots} ,m+1
\label{ex-gamm}
\end{equation}
\end{definition}

For $n=1$ and with notation $u_i \equiv u_i^{(n=1)}$ one derives the following
proposition :
\begin{proposition}
\label{proposition:gamma-beta}
The matrix coefficients of $\Gamma$ satisfy the following equations:    
\begin{eqnarray}
\frac{\partial }{ \partial u_i} \gamma_{j\, k} ({\bf u}, \lambda)  &=& 
\beta_{ji} ({\bf u}) \,
\gamma_{i\, k}  ({\bf u}, \lambda) 
\;\; ;\;\; i \ne j=1,{\ldots},m+1 \label{sigj}\\ 
\sum_{j=1}^{m+1} \frac{\partial }{ \partial u_j}  
\gamma_{i\, k} ({\bf u}, \lambda)  &=& \lambda \,
\gamma_{i\, k}  ({\bf u}, \lambda) )  \;\; ;\;\; i =1,{\ldots},m+1 \label{sumsigj}
\end{eqnarray}
for each $k=1,{\ldots} , m+1$.
\end{proposition}
{\bf Proof}.
Equation (\ref{sumsigj}) follows immediately from relation
\begin{equation}
\frac{\partial }{ \partial u_j^{(n)}} \Gamma = -(\Theta \lambda^n
E_{jj} {\Theta}^{-1})_{-} \Gamma
+ \Theta ({\bf u},\lambda)\lambda^n E_{jj}  D ({\bf u},\lambda) 
= (\Theta \lambda^n E_{jj} {\Theta}^{-1})_{+} \Gamma
\label{sjkgm}
\end{equation}
which can also be rewritten as 
\begin{equation}
\frac{\partial }{ \partial u_j^{(n)}} \Gamma ({\bf u}, \lambda) = 
(\Gamma \lambda^n E_{jj} {\Gamma}^{-1})_{+} \, \Gamma ({\bf u}, \lambda) 
\label{sjkgmgm}
\end{equation}
To proof equation (\ref{sigj}) one can either use (\ref{sjkgmgm}) or 
equations
(\ref{eigenf-higher}) with the identification (\ref{agree}).\hfill{$\square$}
\vskip 10pt \noindent
Equations in Proposition \ref{proposition:gamma-beta} give rise to the
following compatibility conditions :
\begin{corollary}
\label{corollary:compatibi}
The rotation coefficients $\beta_{ij}$ satisfy
\begin{equation}
    \frac{\partial }{ \partial u_k} \beta_{ij} = \beta_{ik} \beta_{kj},
\;\; i \ne k \ne j \;\;\;
\sum_{k=1}^{m+1} \frac{\partial }{ \partial u_k} \beta_{ij}     =0,
\;\; i \ne j \, .
\label{betas-comp}
\end{equation}
\end{corollary}

Similarly, the inverse of the $\Gamma$ matrix 
$\Gamma^{-1}=(\gamma_{ij}^{-1})_{1\le i,j\le m+1}$ is given by :
\begin{equation}
\Gamma^{-1} ({\bf u}, \lambda) 
=  D^{-1} ({\bf u}, \lambda)  \Theta^{-1}  ({\bf u}, \lambda) %\quad ; \quad  D_{ij}^{-1} 
%(s,\lambda) = \delta_{ij}  
%\, e^{-\sum^{\infty}_{n=1} \lambda^n u_j^{(n)}} 
\label{gms-inv}
\end{equation}
or in components 
\begin{equation}
\gamma^{-1}_{k\,i} ({\bf u}, \lambda) = \theta^{-1}_{k\,i} ({\bf u}, \lambda) 
\;
e^{-\sum^{\infty}_{n=1} \lambda^n u_k^{(n)}}
\qquad k,i=1,{\ldots} ,m+1
\label{ex-gamm-star}
\end{equation}
for which we derive the proposition :
\begin{proposition}
\label{gamma-inv-beta}  
The matrix coefficients of $\Gamma^{-1}$ satisfy the following equations:  
\begin{eqnarray}
\frac{\partial }{ \partial u_i} \gamma^{-1}_{k\,j}  ({\bf u}, \lambda)  &=& \gamma^{-1}_{k\,i}  
({\bf u}, \lambda)  \, \beta_{ij} ({\bf u}) 
\;\; ;\;\; i \ne j=1,{\ldots},m+1 \label{sigstarj}\\ 
\sum_{j=1}^{m+1} \frac{\partial }{ \partial u_j}  
\gamma^{-1}_{k\,i}  ({\bf u}, \lambda) &=& -\lambda \,
\gamma^{-1}_{k\,i}  ({\bf u}, \lambda)    \;\; ;\;\; i =1,{\ldots},m+1 \, . 
\label{sumsigstarj}
\end{eqnarray}
\end{proposition}
Furthermore we also have the following :
\begin{proposition}
\label{proposition:bilinear}
The $\Gamma$ and $\Gamma^{-1}$ matrices satisfy the following two bilinear identities 
\begin{eqnarray}
{\rm  Res}_\lambda \left(\Gamma^{-1}({\bf u},\lambda) \Gamma({\bf
u'},\lambda)\right)&=&0
\label{gginv}\\
{\rm  Res}_\lambda \left(\Gamma({\bf u},\lambda)\Gamma^{-1}({\bf u'},\lambda)\right)&=&0
\label{ginvg}
\end{eqnarray}
\end{proposition}
%This follows from (74) and (66) and Taylors formula:
{\bf Proof}$\,$ 
Since
\begin{equation}
\Gamma^{-1} ({\bf u}, \lambda)  \Gamma ({\bf u}, \lambda) 
= \Gamma ({\bf u}, \lambda) \Gamma^{-1} ({\bf u}, \lambda)  
= I \, ,
\label{ggstar}
\end{equation}
one determines that for $k\ge 0$ :
\begin{equation}
{\rm  Res}_\lambda \left(\lambda^k\Gamma^{-1}({\bf u},\lambda)\Gamma({\bf u},\lambda)\right)=0
\label{gginvss}
\end{equation}
and
\begin{equation}
{\rm  Res}_\lambda \left(\lambda^k\Gamma({\bf u},\lambda)\Gamma^{-1}({\bf u},\lambda)\right)=0
\label{ginvgss}
\end{equation}
and hence from relation (\ref{sjkgmgm}) one gets : 
\begin{equation}
{\rm  Res}_\lambda \left(\Gamma^{-1}({\bf u},\lambda)
 \frac{\partial\Gamma({\bf u},\lambda)}{\partial u_j^{(n)}}\right)=0
\label{ginvgpss}
\end{equation}
and a similar formula for equation (\ref{ginvgss}).
The conclusion follows now from the Taylor expansion in $({\bf u'}-{\bf u})$ 
in relations (\ref{gginv})-(\ref{ginvg}).
\hfill{$\square$}
\vskip 10pt \noindent
Notice, now that
\begin{equation}
\exp\left({\sum_{j=1}^{m+1} \sum_{n=1}^{\infty} E^{(n)}_{jj}u^{(n)}_j}\right)
=\sum_{j=1}^{m+1} E_{jj} e^{\sum_{n=1}^{\infty} \lambda^{n} u^{(n)}_j}
= D ({\bf u}, \lambda) 
\label{dfirst}
\end{equation}
where the matrix $D ({\bf u}, \lambda) $ was previously defined in terms of
with its matrix elements
$D_{ij}$ in (\ref{gms}).

Hence the Riemann-Hilbert problem can be recast in the form :
\begin{equation}
D ({\bf u}, \lambda) g
= g_{-}g_{+} = \Theta^{-1} M
\label{new-rh}
\end{equation}
or 
\begin{equation}
\Gamma ({\bf u}, \lambda) g = M ({\bf u}, \lambda)\, , 
\label{gam-rh}
\end{equation}
which provides another proof for that $\Gamma ({\bf u}, \lambda)$ satisfies the evolution eqs.
(\ref{sjkgmgm}) identical to those satisfied by $M ({\bf u}, \lambda)$ in (\ref{umplus}).
In \cite{vdL1} and \cite{LM}  a matrix  similar to $M ({\bf u}, \lambda)$ was defined via 
formula (\ref{gam-rh}) for $\Gamma ({\bf u}, \lambda)$ the 
$n$-component KP wave
function of \cite{KvdL}.
As a corollary of relation (\ref{gam-rh}) we find that the matrix $M ({\bf u}, \lambda)$ 
also satisfies Proposition \ref{proposition:gamma-beta}.

\section{Reduction to the CKP Hierarchy}

In this section we will describe the form of reduction of the integrable
hierarchy presented in the previous section which renders 
the $\beta_{ij}$-coefficients from the Definition \ref{definition:betas}
symmetric.
Clearly, this is accomplished for
\begin{equation}
    \Phi_i\, =\, \Psi_i \quad; \quad  i=1,{\ldots} ,m\, ,   
\label{p=psi}
\end{equation}
which results in the CKP condition \cite{Date81,Loris99}
\begin{equation}
    {\cal L}^{*}\, =\,- {\cal L}
\label{llstar}
\end{equation}
for the pseudo-differential Lax operator from (\ref{lax}).
This condition implies that 
\begin{equation}
\left({\cal L}^{n}\right)^{*}_{+}\, =\,(-)^n \left({\cal L}^{n}\right)_{+}
    \;\; ; \;\;
\left({\cal M}^{(n)}_k\right)^{*}\, =\,(-)^n \left({\cal M}^{(n)}_k\right)
 \label{minn}
\end{equation}
and accordingly the evolution equations (\ref{iso-lax}) and (\ref{add-lax})
are only consistent with condition (\ref{llstar}) for
odd $n$.
Hence, we are interested in the reduced integrable hierarchy which
is obtained from the integrable structure of Section 2 by embedding it
in the CKP hierarchy by imposing condition (\ref{p=psi}) 
and introducing dependence on odd flows $u^{(2k+1)}_i$ 
with $i=1,{\ldots} ,m+1$ only.

It was shown in \cite{Date81}, that all tau-functions corresponding to the CKP hierarchy can be obtained from special KP tau-functions, viz., the ones that satisfy
$$
\tau(t_1,t_2,t_3,t_4,...)=\tau(t_1,-t_2,t_3,-t_4,...)
$$
by putting all $t_{2j}=0$.
{}{}From (\ref{ba-fct}) one easily deduces that the corresponding KP wave functions satisfy
\begin{equation}
\label{g-1}
\psi_{BA}((-)^{i+1}t_i,-\lambda)=\psi_{BA}^*(t,\lambda)
\end{equation}
Now let $W\in Gr^{(0)}(H)$, such that $\psi_W$ satisfies (\ref{g-1}), then 
$$
(f(\lambda),g(-\lambda))=0\ \text{for all }  f,g\in W.
$$
All such elements $W$ form a Grassmann submannifold of $Gr^{(0)}$ which we denote by 
$C(H)$, the Grassmannian of the CKP hierarchy.

To obtain the Darboux--Egoroff system one needs to impose 
condition (\ref{p=psi}) on the Lax operator (\ref{lax}). In the following we arrive at this condition
within the Grassmannian formalism.
Assume now that $W\in C(H)$ which
satisfies  (\ref{g4}). In the discussion in Section \ref{Sat} we have found $m$ independent vectors $v_j(\lambda)\in W'$ and 
$m$ vectors $u_j(\lambda)\in W^\perp$, such that $(v_i(\lambda),u_j(\lambda))=0 $ for $i\ne j$ and 
$(v_i(\lambda),u_i(\lambda))\ne 0$.
Multiply the  $u_j$
with a scalar such that
$$
(v_i(\lambda), u_j(\lambda))=\delta_{ij}.
$$
Since $u_j(\lambda)\in W^\perp$, we observe that $u_j(-\lambda)\in W$.
Express
$v_j(\lambda)=a_j(\lambda)+\lambda b_j(\lambda)$ with 
both $a_j,b_j\in W$ and moreover the $b_j\ne 0$ and linearly independent
(otherwise $W\subset W'$ not codimension $m$).
Then
$$
\begin{aligned}
\delta_{ij}=&(v_j(\lambda),u_i(\lambda))\\
=&(a_j(\lambda)+\lambda b_j(\lambda),u_i(\lambda))\\
=&(\lambda b_j(\lambda),u_i(\lambda)).
\end{aligned}
$$
So we see that we may replace in (\ref{pp}) $v_j(\lambda)$ by $\lambda b_j(\lambda)$, this does not change 
the definition of $\Psi_j$ since $(\psi_W^*(t,\lambda),a_j(\lambda))=0$.
Now notice that for $b,c\in W$
\begin{equation}
\label{gg1}
\begin{aligned}
(\lambda b(\lambda), c(-\lambda))=&
-(-\lambda b(-\lambda), c(\lambda))\\
=&
(\lambda b(-\lambda), c(\lambda))\\
=&
 (\lambda c(\lambda), b(-\lambda)),
\end{aligned}
\end{equation} 
hence $(\lambda u_j(-\lambda), b_j(-\lambda))\ne 0$ and therefore
$\lambda  u_j(-\lambda)\not\in W$ for all $j$.
Write 
$$
\lambda  u_j(-\lambda)=w_j(\lambda) +\sum_{\ell=1}^m B_{j\ell}\lambda b_\ell(\lambda),
$$
with $w_j\in W$, then 
clearly,
$$
\left(\psi_W(t,\lambda),\lambda u_j(\lambda)\right)=\left(\psi_W(t,\lambda),\sum_{\ell=1}^m B_{j\ell} \lambda b_\ell(-\lambda)\right),
$$
since $w_j(-\lambda)\in W^\perp$.
So we can replace $u_j(\lambda)$ in (\ref{pp}) by 
$$
\sum_{\ell=1}^m B_{j\ell} b_\ell(-\lambda),
$$
this does not change the eigenfunctions.
We calculate
$$
\begin{aligned}
\delta_{ij}&=(u_j(\lambda), v_i(\lambda))\\
&=(u_j(\lambda),a_i(\lambda)+\lambda b_i(\lambda))\\
&=(u_j(\lambda),\lambda b_i(\lambda))\\
&=(\lambda u_j(\lambda), b_i(\lambda))\\[2mm]
&=
\left(-w_j(-\lambda) +\sum_{\ell=1}^m B_{j\ell} \lambda b_\ell(-\lambda), b_i(\lambda)\right)\\[2mm]
&=
\left(\sum_{\ell=1}^m B_{j\ell} \lambda b_\ell(-\lambda), b_i(\lambda
)\right)\\
&=
\sum_{\ell=1}^m B_{j\ell} (\lambda b_\ell(-\lambda), b_i(\lambda))
\end{aligned}
$$
In other words, let $B=(B_{ij})_{1\le i,j\le m}$,
then 
$$
B ( \lambda b_\ell(-\lambda), b_i(\lambda)))_{1\le \ell ,i\le m}= I_m
$$
and thus 
$$
( \lambda b_\ell(-\lambda), b_i(\lambda)))_{1\le \ell ,i\le m}
$$
is invertible.
Now denote
$$
A=( (\lambda b_i(-\lambda), b_j(\lambda) )_{1\le i,j\le m}=( (\lambda b_j(\lambda),b_i(-\lambda)) )_{1\le i,j\le m}.
$$
Substituting $b=b_j,c=b_i$ in (\ref{gg1}) one sees
that
$$(\lambda b_i(-\lambda), b_j(\lambda))= (\lambda b_j(-\lambda), b_i(\lambda))
$$
so $A$ is symmetric (but not necessarily real). Hence one can find a new basis
$$
h_i(\lambda)\in \sum_{j=1}^m \mathbb{C}b_j(\lambda),
$$
 $h_i\in W$
such that 
\[
(h_i(-\lambda), h_j(\lambda))=\delta_{ij}.
\]
Choosing this basis instead of the original one, our Lax operator has the form
$$
{\cal L}_W= \partial_x+\sum_{i=1}^m c_i(\psi_W(t,\lambda),\lambda h_i(-\lambda))\partial_x^{-1} (\psi_W^*(t,\lambda), \lambda h_i(\lambda)),\ \text{ for certain }c_i\in\mathbb{C}.
$$
Now multiplying $h_i$ with $\sqrt{c_i}$, we obtain
\begin{proposition}
\label{prop laxform}
Let $W\in C(H)$ satisfying (\ref{g4}), then there exist $m$ independent functions $h_i(\lambda)\in W$ with 
\begin{equation}
\label{h}
(h_i(-\lambda), h_j(\lambda))=\delta_{ij}c_i,\quad 0\ne c_i\in \mathbb{C},
\end{equation}
such that
\begin{equation}
\label{glax}
\begin{aligned}
{\cal L}_W&= \partial_x+\sum_{i=1}^m \Phi_i\partial_x^{-1}\Psi_i,\quad\text{with}\\
\Phi_i(t)&=(\psi_W(t,\lambda),\lambda h_i(-\lambda)),\\
\Psi_i(t)&= (\psi_W^*(t,\lambda), \lambda h_i(\lambda))\\
\ &=(\psi_W((-)^{j+1}t_j,-\lambda), \lambda h_i(\lambda))\\
\ &=(\psi_W((-)^{j+1}t_j,\lambda), \lambda h_i(-\lambda)),
\end{aligned}
\end{equation}
\end{proposition}
So we have constructed (adjoint) eigenfunctions $\Phi_i$ and $\Psi_i$ that satisfy
\begin{equation}
\label{symmpp}
\Phi_i(t)|_{t_{2j}=0\ \text{for all }j\ge 1}=\Psi_i(t)|_{t_{2j}=0\ \text{for all }j\ge 1} .
\end{equation}

{\it The CKP or reduced hierarchy  is now obtained
by putting not only all $t_{2j}=0$, but also all $u_{k}^{(2j)}=0$ for all $1\le k\le m+1$, $j=1,2,3,\ldots$ (cf. Definition \ref{definition:sato-add-flows}). We will assume this to hold from now on.}
As a consequence  of the above we find that  
\begin{proposition}
\label{proposition:CKPth} 
For the reduced hierachy the rotation coefficients $ \beta_{ij}({\bf u})$
satisfy the Darboux--Egoroff system (\ref{D3}), (\ref{D5}). One also has
\begin{eqnarray}
\Theta^{-1} ( {\bf u}, \lambda ) &=&     \Theta^T ( {\bf u}, -\lambda )
\quad ;\quad 
\Gamma^{-1} ( {\bf u}, \lambda ) =     \Gamma^T  ( {\bf u}, -\lambda )
\label{tggt}\\
{\rm  Res}_\lambda \left(\Gamma({\bf u},\lambda)\Gamma^T ({\bf u'},-\lambda)\right)&=&0
\;\; ; \;\; {\rm  Res}_\lambda \left(\Gamma^T ({\bf u},-\lambda)\Gamma({\bf u'},\lambda)\right)=0
\label{ginvg-CKP}
\end{eqnarray}
\end{proposition}

\section{Construction of the WDVV Prepotential}
In this section we want to construct the WDVV prepotential $F$, from Section \ref{Darb}, in terms of the data of the reduced hierarchy, which is obtained by putting all $u_{k}^{(2j)}=0$ for all $1\le k\le m+1$, $j=1,2,3,\ldots$. Recall from Proposition \ref{proposition:CKPth} that the rotation coefficients $\beta_{ij}$ satisfy the Darboux--Egoroff system (\ref{D3}), (\ref{D5}).
\begin{definition}
Choose some fixed ${\bf u}'$ such that ${\bf u}'\ne {\bf u}$, for which
\begin{equation}
{\rm  Res}_\lambda \left(\lambda^{-1}
\Gamma({\bf u},\lambda)\Gamma^T ({\bf u'},-\lambda)\right)\ne 0,
\label{lgg}
\end{equation}
then define
\label{definition:goodpsi}
\begin{equation}
\Psi({\bf u},\lambda)= \Gamma({\bf u},\lambda)\, \Gamma^T ({\bf u'},-\lambda)\, .
\label{psiggll}
\end{equation}
\end{definition}
It is always possible to find such a ${\bf u}'$  due to
\begin{equation}
\Gamma({\bf u},\lambda)\Gamma^T ({\bf u},-\lambda) = I.
\label{lggii}
\end{equation}

In view of Proposition \ref{proposition:gamma-beta} the following results hold
\begin{proposition}
\label{proposition:CKPpsi}
The matrix $\Psi({\bf u},\lambda)$ is a positive power series in $\lambda$, whose matrix 
coefficients satisfy the equations \ref{D10}, i.e.,
\begin{eqnarray}
\frac{\partial }{ \partial u_i} \Psi_{j\, k} ({\bf u}, \lambda)  &=& 
\beta_{ji} ({\bf u}) \,
\Psi_{i\, k}  ({\bf u}, \lambda) 
\;\; ;\;\; i \ne j=1,{\ldots},m+1 \label{pssigj}\\ 
\sum_{j=1}^{m+1} \frac{\partial }{ \partial u_j}  
\Psi_{i\, k} ({\bf u}, \lambda)  &=& \lambda \,
\Psi_{i\, k}  ({\bf u}, \lambda) )  \;\; ;\;\; i =1,{\ldots},m+1 \label{sumpssigj}
\end{eqnarray}
for each $k=1,{\ldots} , m+1$.
\end{proposition}
{\bf Proof} The equations (\ref{pssigj}) and (\ref{sumpssigj}) follow immediately from Proposition \ref{proposition:gamma-beta}. Next observe from 
(\ref{ginvg-CKP}) that
\begin{equation}
\label{bla}
{\rm Res}_\lambda \lambda^j\Psi({\bf u},\lambda)=0
\end{equation} for $j=0$.
Then using (\ref{sumpssigj}), we see that (\ref{bla}) even holds for all $j\ge 0$, i.e., $\Psi({\bf u},\lambda)$ is a power series in $\lambda$.\hfill{$\square$}
\vskip 10pt
\noindent
Using formula (\ref{gam-rh})
we also have that 
\begin{equation}
\Psi ({\bf u}, \lambda) = M ({\bf u}, \lambda) 
M^{-1} ({\bf u'}, \lambda) 
\label{psi-m}
\end{equation}
So that $\Psi ({\bf u}, \lambda)$ can be expressed by $\Gamma$ from $G_{-}$
as well as by $M$ from $G_{+}$.
This provides a different way to observe that $
\Psi ({\bf u}, \lambda) $ is a positive expansion in
$\lambda$.

As a consequence of Proposition  \ref{proposition:CKPpsi} the constant coefficients $\Psi_{i\, j}  ({\bf u}, 0)$ of the matrix
coefficients $\Psi_{i\, j}  ({\bf u}, \lambda)$ satisfy (\ref{D11}) and therefore 
provide the data of Proposition \ref{Prop1}. 
In particular the Lam\'e coefficients are equal to
\begin{equation}
\label{WDVV-lame}
h_i({\bf u})=\Psi_{i\, 1}  ({\bf u}, 0).
\end{equation}
The higher times $u^{(2k+1)}_j$ for $k>0$ play the role of parameters.
Also, due to (\ref{ggstar}) it holds that :
\begin{equation}
\Psi({\bf u},\lambda)\Psi^T({\bf u},-\lambda)=\Psi^T({\bf u},-\lambda)\Psi({\bf u},\lambda)=I.
\label{pspst}
\end{equation}
{}{}From this we deduce that in fact $\eta_{\alpha\beta}=\delta_{\alpha\beta}$.
Now define as in Definition \ref{def1}
\begin{definition}
\label{def1again}
The matrix $\Xi'$ and row-vector $\Xi$ are defined as:
\[
\begin{aligned}
\Xi'({\bf u},\lambda)&=\left(\Psi_{i1}({\bf u},0)\Psi_{ij}({\bf u},\lambda)\right)_{ij},\\[2mm]
\Xi({\bf u},\lambda)&=\left(\Xi({\bf u},\lambda)_j\right)_j=\lambda^{-1}\left(\sum_{i=1}^{m+1}\Psi_{i1}({\bf u},0)\Psi_{ij}({\bf u},\lambda)\right)_{j},\\
\Xi({\bf u},\lambda)_j&=\delta_{j1}\lambda^{-1}+x^j({\bf u})+\sum_{i=1}^\infty \xi_i^j({\bf u})\lambda^i.
\end{aligned}
\]
\end{definition}
Then these $\Xi'$ and $\Xi$ satisfy (\ref{f5.10})-(\ref{theta-prod}) and hence we obtain (see Proposition \ref{Prop1} and Theorem \ref{theor Akm})
\begin{theorem}
For the reduced hierarchy the rotation coefficients $\beta_{ij}({\bf u})$ satisfy the Darboux--Egoroff system (\ref{D3}), (\ref{D5}).
On the domain
$u_i^{(1)}\ne u_j^{(1)}$ and $\Psi_{11}({\bf u},0)\Psi_{21}({\bf u},0)\cdots\Psi_{n1}({\bf u},0)\ne 0$, one has
\[
\begin{aligned}
h_i({\bf u})&=\Psi_{i1}({\bf u},0),\\[2mm]
\eta_{\alpha\beta}&=\delta_{\alpha,\beta},\\[2mm]
x^\alpha({\bf u})&=\sum_{i=1}^{m+1}\Psi_{i1}({\bf u},0){\rm Res}_\lambda \left(\lambda^{-2}\Psi_{i\alpha}({\bf u},\lambda)\right),\\
c_{\alpha\beta}^{\gamma}({\bf u})&=c_{\alpha\beta\gamma}({\bf u})=\sum_{i=1}^{m+1}
\frac{\Psi_{i\alpha}({\bf u},0)\Psi_{i\beta}({\bf u},0)\Psi_{i\gamma}({\bf u},0)}{\Psi_{i1}({\bf u},0)}.
\end{aligned}
\]
The function $F({\bf u})$ defined by
\[
F({\bf u})=-\frac{1}{2}\xi_2^1({\bf u})+\frac{1}{2}\sum_{i=1}^{m+1}
x^i({\bf u})\xi_1^i({\bf u}),
\]
with
\[
\xi_i^j({\bf u})=\sum_{k=1}^{m+1}\Psi_{k1}({\bf u},0){\rm Res}_\lambda \left(\lambda^{-2-i}\Psi_{kj}({\bf u},\lambda)\right)
\]
satisfies the WDVV-equations (\ref{D9}).

\end{theorem}

\vskip 10pt \noindent
{\bf Acknowledgements} \\
H.A. is partially supported by NSF (PHY-9820663).

%==

\end{document}